\documentclass[preprint,11pt]{aastex}
\usepackage{epsfig}
\usepackage{rotating}      
\tightenlines

\newcommand	\alphamax	{{\pi/2}}

\newcommand	\Arcmin	{\,{\rm arcmin}}

\newcommand	\beq	{\begin{equation}}
\newcommand	\Chandra	{{\it Chandra}}
\newcommand	\cm	{\,{\rm cm}}
\newcommand	\ct	{\,{\rm ct}}
\newcommand	\days	{\,{\rm d}}

\newcommand	\eeq	{\end{equation}}
\newcommand	\erg	{\,{\rm ergs}}

\newcommand	\eV	{\,{\rm eV}}

\newcommand	\g		{\,{\rm g}}
\newcommand	\gtsim	{\gtrsim}		 

\newcommand	\keV	{\,{\rm keV}}
\newcommand	\K		{\,{\rm K}}
\newcommand	\kms	{\,{\rm km~s}^{-1}}
\newcommand	\kpc	{\,{\rm kpc}}

\newcommand	\ltsim	{\lesssim}		 

\newcommand	\pc	{\,{\rm pc}}

\newcommand	\PSPC	{{\it PSPC}}
\newcommand	\rmH	{{\rm H}}
\newcommand	\ROSAT	{{\it ROSAT}}
\newcommand	\Rwd	{R_{\rm wd}}
\newcommand	\s		{\,{\rm s}}

\newcommand	\sr	{\,{\rm sr}}
\newcommand	\thetascat {{\Theta_s}}
\newcommand	\Twd	{T_{\rm wd}}

\newcommand	\XMM	{{\it XMM}}

\newlength{\figwidth}
\countdef\decade=200
\advance\decade by \year
\countdef\hours=201
\advance\hours by \time
\divide\hours by 60
\countdef\mins=202
\advance\mins by \hours
\multiply\mins by 60
\multiply\hours by 100
\countdef\miltime=203
\advance\miltime by \hours
\advance\miltime by \time
\advance\miltime by -\mins

\begin{document}

\title{
        \vspace*{-3.0em}
        {\normalsize\rm submitted to {\it The Astrophysical Journal}}\\ 
        \vspace*{1.0em}
	The Scattered X-Ray Halo Around Nova Cygni 1992:\\
	Testing a Model for Interstellar Dust
	}

\author{B. T. Draine and Jonathan C. Tan
	}
\affil{Princeton University Observatory, Peyton Hall, Princeton,
NJ 08544; {\tt draine@astro.princeton.edu, jt@astro.princeton.edu}}

\begin{abstract}
We use published \ROSAT\ observations of the X-ray Nova V1974 Cygni 1992 
to test a model for interstellar dust,
consisting of a mixture of carbonaceous grains and silicate grains.
The time-dependent X-ray emission from the 
nova is modelled as the sum of emission from a O-Ne white dwarf plus
a thermal plasma, and X-ray scattering is calculated for a dust mixture
with a realistic size distribution.
Model results are compared 
with the scattered X-ray halos measured by \ROSAT\ at 9 different
epochs, including the early period of rising X-ray emission,
the ``plateau'' phase of steady emission, and the
X-ray decline at late times.
We find that the observed X-ray halos appear to be consistent with
the halos calculated for the size distribution of Weingartner \& Draine which
reproduces the Milky Way extinction with $R_V=3.1$, provided
that the reddening to the nova is $E(B-V)\approx 0.20$,
consistent with 
$E(B-V)\approx 0.19$ 
inferred from the late-time Balmer
decrement.
The time delay of the scattered halo relative to the direct flux from
the nova is clearly detected.

Models with smoothly-distributed dust give good overall agreement with
the observed scattering halo, but tend to produce somewhat more scattering than
observed at 200--300\arcsec, and insufficient scattering at 50--100\arcsec.
While an additional population of large grains can 
increase the
scattered intensity at 50--100\arcsec,
this could also be achieved by having $\sim$30\% of the dust in a cloud
at a distance from us equal to $\sim$95\% of the distance to the nova.
Such a model also improves agreement with the data at larger angles,
and illustrates the sensitivity of X-ray scattering halos to the location
of the dust.
The observations therefore do
not require a population of micron-sized dust grains.

Future observations by \Chandra\ 
and {\it XMM-Newton}
of X-ray scattering halos around extragalactic
point sources can provide more stringent tests of interstellar dust
models.

\end{abstract}


\section{Introduction
	\label{sec:intro}}

Interstellar grains scatter X-rays through small scattering angles,
and as a result distant X-ray point sources appear to be surrounded
by a diffuse ``halo'' of scattered X-rays 
(Overbeck 1965; Martin 1970; Hayakawa 1973).
The angular structure and absolute intensity of
these scattered halos can be measured using
imaging X-ray telescopes,
thus providing a test for interstellar grain 
models (Catura 1983; Mauche \& Gorenstein 1986; Mitsuda et al.\ 1990;
Mathis \& Lee 1991; 
Clark et al.\ 1994;
Woo et al.\ 1994; Mathis et al.\ 1995; 
Predehl \& Klose 1996; Smith \& Dwek 1998;
Witt et al.\ 2001;
Smith et al.\ 2002).
Given an accurate model for the dust grain size distribution and its
scattering properties, observations of scattering halos can also constrain
the spatial distribution of dust towards a source and the distance to a
source, particularly if the emission is 
time variable (Tr\"umper \& Sch\"onfelder 1973; Predehl et al.\ 2000).

Nova V1974 Cygni 1992, a bright X-ray nova, was
observed extensively by the imaging X-ray telescope on \ROSAT,
resulting in the best extant data set for 
studies of the X-ray scattering properties of dust
(Krautter et al.\ 1996).
Mathis et al.\ (1995) compared model calculations to
the X-ray halo observed 291 days
after optical maximum, and argued that the angular structure of the
observed X-ray halo favored
a grain model based on highly porous grains.
Smith \& Dwek (1998) disagreed with this conclusion, arguing that
the halo around Nova Cygni 1992 did not require porous grains, but was
in fact consistent with the scattering expected from a mixture of nonporous
silicate and carbon grains.
More recently,
Witt, Smith, \& Dwek (2001) reached a different conclusion,
arguing that the observed X-ray halo around Nova Cygni 1992 requires that the
size distribution of interstellar dust grains extend to radii 
$a\geq2.0\micron$, with $\gtsim$40\% of the dust mass in grains with
radii $a > 0.5\micron$.

Weingartner \& Draine (2001) and Li \& Draine (2001) have recently
put forward a physical dust model which is in quantitative
agreement with 
the wavelength-dependent extinction of starlight as well as
the observed spectrum of infrared emission from interstellar dust.
The model consists of a mixture of carbonaceous grains (including ultrasmall
grains with the properties of polycyclic aromatic hydrocarbon molecules)
and amorphous silicate grains.
By appropriate adjustment of the size distribution, the model can
reproduce the extinction in different regions of the Milky Way, 
and in the Large and Small Magellanic Clouds.
Li \& Draine (2002) show that the model is also consistent with
the observed infrared emission from the Small Magellanic Cloud.
Here we use the observed X-ray halo around Nova Cygni 1992 to test
this dust grain model.
Since the sightline toward Nova Cygni 1992 is presumably typical
diffuse interstellar medium, we use the 
Weingartner \& Draine (2001, hereafter WD01)
size distribution for Milky Way dust with $R_V=3.1$.

In \S\ref{sec:distance} we review estimates of the distance to
and the gas and dust toward Nova Cygni 1992.  
An empirical model for the
X-ray emission from the nova is described in \S\ref{sec:spectrum}, with
the emission modelled as the sum of emission from a hot thermal plasma
plus a white dwarf photosphere with varying temperature and radius.
The methodology for calculation of X-ray scattering by dust is presented
in \S\ref{sec:scatter}, including multiple scattering, the effects of
time delay, and the calculation of dust scattering cross sections.

Our results are presented in \S\ref{sec:results}.
We test our model using
observations from 9 different epochs, at radii out to 2000\arcsec.
We show that
the WD01 model is in quite good agreement with the observations if the
dust is assumed to follow an exponential density law and the nova
is at a distance of $\sim$2.1~kpc.
The time delay of the halo relative to the nova is clearly visible at
late times when the nova is in decline.
We discuss the uncertainties associated with possible clumping of the
dust into clouds along the line-of-sight, and show that agreement with
the observed halos can be improved if $\sim30\%$ of the dust is concentrated
in a cloud $\sim100\pc$ from the nova.
We conclude that the WD01 dust model is consistent with
the observed X-ray halo toward Nova Cygni 1992, and
a population of large dust grains is not required.

The distance estimate to the nova depends somewhat on the assumed
density distribution of dust, and is considered in more detail in a
separate paper (Draine \& Tan 2003).

\section{X-Ray Observations\label{sec:XRayObs}}

\ROSAT\ \PSPC\ images of Nova Cygni 1992 for days\footnote{%
	Day ``258'' is the sum of observations on days 255 and 259.
	Day ``650'' is the sum of observations on days 647, 648, 652,
	and 653.%
	}
258, 261, 291, 434, 511, 612, 624, 635, and 650 after optical maximum
were extracted from the \ROSAT\ archive and analysed using the ESAS
software package (Snowden et al.\ 1994). This corrects for the effects
of the nonuniform detector response (particularly the shadowing by
struts around 1200\arcsec) and vignetting, and excludes periods of high
solar activity, allowing fairly accurate X-ray intensity profiles to
be determined out to $\sim 2000-3000\arcsec$, particularly when the
nova was bright.  From the observation at day 650, which has the
deepest exposure and the weakest nova halo, we identified the nine
brightest background point sources.  Excluding these sources,\footnote{%
	Only 1RXS J202742.6+522920 ($\theta=1630\arcsec$) with 0.09~ct/s and
	1RXS J202742.4+523621 ($\theta=1510\arcsec$) 
	with 0.035~ct/s make significant 
	contributions to the halos.%
	}
the azimuthally-averaged intensities and their statistical uncertainties
were calculated for all annuli around the nova.

The diffuse background was estimated for each epoch by
averaging over an annulus from 2800\arcsec\ to 3200\arcsec, where the
angular dependence of intensity was seen to be flat. Backgrounds
ranged from $5.2\times10^{-4}\:{\rm ct} \s^{-1} \Arcmin^{-2}$ (day
261) to $9.9\times 10^{-4}\:{\rm ct} \s^{-1} \Arcmin^{-2}$ (day 624).
This scatter is most likely due to uncertainties in estimating
intrumental backgrounds in the ESAS data reduction process.
Background subtraction was therefore done separately for each epoch so
that these systematic uncertainties would have a reduced impact on the
determination of the intensity of the nova's halo. 
Note, however, this method of
evaluating the background forces the derived halos to artificially go
to zero at 3000\arcsec. 
In Table \ref{tab:count_rates} 
we list the background-subtracted count rates in selected
annuli; we do not go beyond $2040\arcsec$
because uncertainties in the background correction dominate at larger
angles.
The error estimates in Table {\ref{tab:count_rates}} include only
statistical errors plus an estimated $\pm10\%$ error in 
the background estimate for each epoch.
\begin{table}[h]
\begin{center}
\caption{X-Ray Lightcurve
	\label{tab:count_rates}}
{\footnotesize
\begin{tabular}{
		c c c c c c
		}
\tableline\tableline\\
epoch	& $\dot{N}(<\!50\arcsec)$
		& $\dot{N}(50\!-\!100\arcsec)$
			& $\dot{N}(100\!-\!300\arcsec)$
				&$\dot{N}(300\!-\!960\arcsec)$
				&$\dot{N}(960\!-\!2040\arcsec)$\\
(day)	& ct/s	& ct/s	& ct/s	& ct/s	& ct/s	\\
\tableline
258	&$11.47\pm.04$
		&$0.522\pm.009$
			&$0.574\pm.011$
				&$0.93\pm.05$
					&$0.38\pm0.17$\\
261	&$15.10\pm.07$
		&$0.628\pm.015$
			&$0.619\pm.016$
				&$1.06\pm.04$
					&$0.60\pm0.15$\\
291	&$29.42\pm0.11$
		&$1.12\pm.02$
			&$1.23\pm.02$
				&$2.14\pm.07$
					&$1.1\pm0.3$\\
434	&$68.09\pm0.15$
		&$2.26\pm.03$
			&$2.68\pm.03$
				&$4.99\pm.07$
					&$3.0\pm0.2$\\
511	&$66.41\pm0.17$
		&$2.45\pm.03$
			&$2.74\pm.04$
				&$5.14\pm.08$
					&$3.3\pm0.2$\\
612	&$3.68\pm.02$
		&$0.200\pm.005$
			&$0.309\pm.009$
				&$0.81\pm.06$
					&$0.8\pm0.2$\\
624	&$1.37\pm.02$
		&$.093\pm.005$
			&$0.168\pm.010$
				&$0.53\pm.07$
					&$0.6\pm0.3$\\
635	&$0.589\pm.011$
		&$.049\pm.003$
			&$.091\pm.008$
				&$0.31\pm.07$
					&$0.2_{-0.2}^{+0.3}$\\
650	&$0.225\pm.004$
		&$.0251\pm.0015$
			&$.041\pm.006$
				&$0.18\pm.06$
					&$0.13_{-0.13}^{+0.21}$\\
\tableline
\end{tabular}
}
\end{center}
\end{table}

Suppose that the point spread function (psf) 
has a fraction $\gamma(\theta)$ of the total counts within an angle $\theta$.
At the median photon energy $E=480\eV$ of the detected photons,
the \ROSAT\ psf has 
$\gamma(50\arcsec)\approx0.9800$ and 
$\gamma(100\arcsec)\approx0.9869$
(Boese 2000).
Suppose that $g(\theta)$ is the fraction of the halo photons interior to
$\theta$.  Then (neglecting the effect of the psf on the scattered photons)
the total point source count rate is
\beq
\dot{N}_{\rm ps} = 
\frac{\dot{N}(0-50\arcsec) -\beta\dot{N}(50\arcsec-100\arcsec)}
	{\gamma(50\arcsec)-
	\left[\gamma(100\arcsec)-\gamma(50\arcsec)\right]\beta} ~~~,
\eeq
\beq
\beta = \frac{g(50\arcsec)}{g(100\arcsec)-g(50\arcsec)} ~~~.
\eeq
A uniform surface-brightness halo would have $\beta=1/3$, but 
halos for continuously-distributed dust have $g(\theta)\propto\theta$
for $\theta\rightarrow 0$ (Draine 2003b), corresponding to $\beta=1$ if this
behavior applied out to $100\arcsec$.
We take $\beta\approx 0.9$ as providing a good approximation to the
actual models.

Table \ref{tab:ptsrc_and_halo} presents our derived point source
count rate 
$\dot{N}_{\rm ps}$, the estimated halo count rate
$\dot{N}_{\rm halo}$ (including only halo angles $\theta < 2040\arcsec$), 
and the
ratio $\dot{N}_{\rm halo}/\dot{N}_{\rm ps}$
of halo counts to point source counts.
Note that $\dot{N}_{\rm halo}/\dot{N}_{\rm ps}$ becomes quite large
at late times -- this is because the point source count rate is
declining rapidly, but the longer light travel time
means that the halo photons were emitted from the nova
at an earlier (more luminous) time than the unscattered photons.
\begin{table}[h]
\begin{center}
\caption{Point Source and Halo Components
	\label{tab:ptsrc_and_halo}}
{\footnotesize
\begin{tabular}{
		c c c c
		}
\tableline\tableline
epoch	& $\dot{N}_{ps}$
		& $\dot{N}_{halo}$
			&$\dot{N}_{halo}/\dot{N}_{ps}$\\
(days)	& ct/s		& ct/s		\\
\tableline
258	&$11.30\pm.05$	&$2.6\pm0.2$	&$0.23\pm.02$\\
261	&$14.93\pm.08$	&$3.1\pm0.2$	&$0.206\pm.013$\\
291	&$29.18\pm0.12$	&$5.8\pm0.3$	&$0.200\pm.011$\\
434	&$67.84\pm0.16$	&$13.2\pm0.3$	&$0.195\pm.005$\\
511	&$65.93\pm0.18$	&$14.1\pm0.3$	&$0.213\pm.005$\\
612	&$3.59\pm.02$	&$2.2\pm0.3$	&$0.62\pm.08$\\
624	&$1.32\pm.02$	&$1.4\pm0.4$	&$1.1\pm0.3$\\
635	&$0.560\pm.012$	&$0.7\pm0.3$	&$1.2\pm0.6$\\
650	&$0.208\pm.004$	&$0.40\pm0.28$	&$1.9\pm1.3$\\
\tableline
\end{tabular}
}
\end{center}
\end{table}

At times $t \leq 511$, the ratio of halo to point source is
approximately constant.  Because day 511 appears to have been
preceded by $\sim$100~days
of nearly constant X-ray luminosity,
the observed halo at day 511 can be interpreted as due to
steady illumination, and we can estimate
the dust scattering optical depth to be
\beq
\tau_{\rm sca}(\theta < 2040\arcsec) \approx \ln \left[1+\left(\dot{N}_{\rm halo}/\dot{N}_{\rm ps}\right)_{511}\right]
\approx 0.193\pm 0.004
\eeq
at a characteristic energy $E\approx 480\eV$.
Based on the modelling discussed below, we estimate that 
$\sim$91.6$\pm2\%$ of the
halo is at $\theta < 2040\arcsec$; the total X-ray scattering optical depth
toward nova Cygni 1992 is therefore $\tau_{\rm sca} \approx 0.211\pm.006$

\section{Distance and Gas Distribution\label{sec:distance}}

Nova Cygni 1992 ($\alpha_{2000}=20^h30^m31.^s76$,
$\delta_{2000}=+52^o37^{\prime}52.^{\prime\prime}9$; Austin et al.\ 1996) has
Galactic coordinates $l=89.1^\circ$, $b=7.82^\circ$ and is
located at a height
$Z_{\star}=D\sin b= 272 (D/2\kpc)\pc$ above the Galactic plane.
The distance $D$ has been controversial, with recent estimates
$3.2\pm0.5\kpc$ (Paresce et al.\ 1994),
$D=2.1\pm0.7\kpc$ (Austin et al.\ 1996),
$1.8\pm0.1\kpc$ (Chochol et al.\ 1997), and
$D=2.6\pm0.25\kpc$ (Balman et al.\ 1998).

To calculate X-ray scattering by dust, we require a model for the
spatial distribution of dust between us and the nova.
We take the Sun to be located at the Galactic midplane, $z=0$.
We model the distribution of interstellar gas by
an exponential distribution:
\beq
\label{eq:exp}
n_{\rm H}(z) = \frac{N_{\rm H}^{\rm ISM}(\infty)\sin b}{h_e} \exp(-z/h_e), ~~~ 
h_e = \frac{z_{1/2}}{\ln2} ~~~,
\eeq
\beq
\label{eq:N_H^ISM_exp}
N_{\rm H}^{\rm ISM}(D) = N_{\rm H}^{\rm ISM}(\infty)
\left[1-\exp(-D\sin b/h_e)\right]
~~~,
\eeq
where $z$ is the height above the Galactic plane,  and
$z_{1/2}$ is the height where the density is 50\% of the maximum
value.

H~I 21 cm emission maps of this region (Hartmann \& Burton 1997)
indicate a total
atomic H column density $N({\rm H~I})=2.2\times10^{21}\cm^{-2}$ 
within $\sim$5~kpc 
(integration performed from $-50\kms<v_{\rm lsr}<+50\kms$; 
a weak component at $-150\kms < v_{\rm lsr} < -50\kms$ is excluded),
if the emission is optically thin.
From the COBE and DIRBE far-infrared maps, Schlegel et al.\ (1998) estimate
a total dust column with $E(B-V)=0.412$~mag;
with the local ratio
$N_{\rm H}/E(B-V)=5.8\times10^{21}\cm^{-2}$ (Bohlin, Savage, \& Drake 1978)
this
corresponds to
$N_{\rm H}^{\rm ISM}(\infty)=2.39\times10^{21}\cm^{-2}$, which we adopt as our
best estimate in the direction of the
nova.

Based on studies of the vertical distribution of the gas (see Binney
\& Merrifield 1998, Fig.\ 9.25)
we adopt a half-density height $z_{1/2}=300$~pc for the
gas at Galactocentric radius $R_0 < R < 1.03R_0$.
This gives a midplane density 
$n_{\rm H}(0)=N_{\rm H}^{\rm ISM}(\infty)\sin b/h_e =0.24\cm^{-3}$.
We will also consider $z_{1/2}=250\pc$ and $350\pc$ for comparison.

Given an adopted density law, we treat the distance $D$ to the
nova as an adjustable parameter which determines the
column of dust and gas between us
and the nova, which in turn determines the strength of the scattering halo
relative to the X-ray point source.
By comparing models with different $D$, we will determine what
column density $N_{\rm H}^{\rm ISM}(D)$ best reproduces the observed
strength of the scattering halo.

\subsection{Attenuation by Gas and Dust}

Radiation from the nova is attenuated 
by gas and dust along the line of sight.
From the \ROSAT\ data, 
Balman et al.\ (1998) find that the observed flux from the nova 
is consistent with emission from hot plasma plus
a white dwarf photosphere, attenuated by
absorption by interstellar gas with column density
$N_{\rm H}\approx2.1\times10^{21}\cm^{-2}$ at late times ($t \gtsim 255\days$),
but with additional absorption at earlier times.

Our objective is simply to find an empirical description which reproduces
the observed count rate and energy spectrum of 
unscattered photons arriving at the Earth.
To this end we adopt the parameters estimated by Balman et al.
The results of Balman et al.\ appear to be consistent with
time-dependent absorption by
\begin{eqnarray}
N_{\rm H}^{\rm abs}(t) &=&  2.1\times 10^{21}
	\left(\frac{255\days}{t}\right)^2\cm^{-2} 
	~~~{\rm for}~t < 255\days\\
	&=& 2.1\times10^{21}\cm^{-2} ~~~{\rm for}~t > 255\days
~~.
\end{eqnarray}
The decline of $N_{\rm H}^{\rm abs}$ with time is presumed to be
due to the combined effects of expansion and ionization of
gas associated with the nova.
We will assume that the absorption is due to the interstellar
contribution $N_{\rm H}^{\rm ISM}$ 
[eq. (\ref{eq:N_H^ISM_exp})]
plus a time-variable
\beq
\Delta N_{\rm H}(t)= N_{\rm H}^{\rm abs}(t) - N_{\rm H}^{\rm ISM} 
\eeq
contributed by gas associated with the nova.\footnote{%
	Note that for all the nova distances and density distributions 
	we consider, $N_{\rm H}^{\rm ISM}<2.1\times 10^{21}\:{\rm cm^{-2}}$, 
	so that $\Delta N_{\rm H}>0$.%
	}

\begin{figure}[h]
\begin{center}
\epsfig{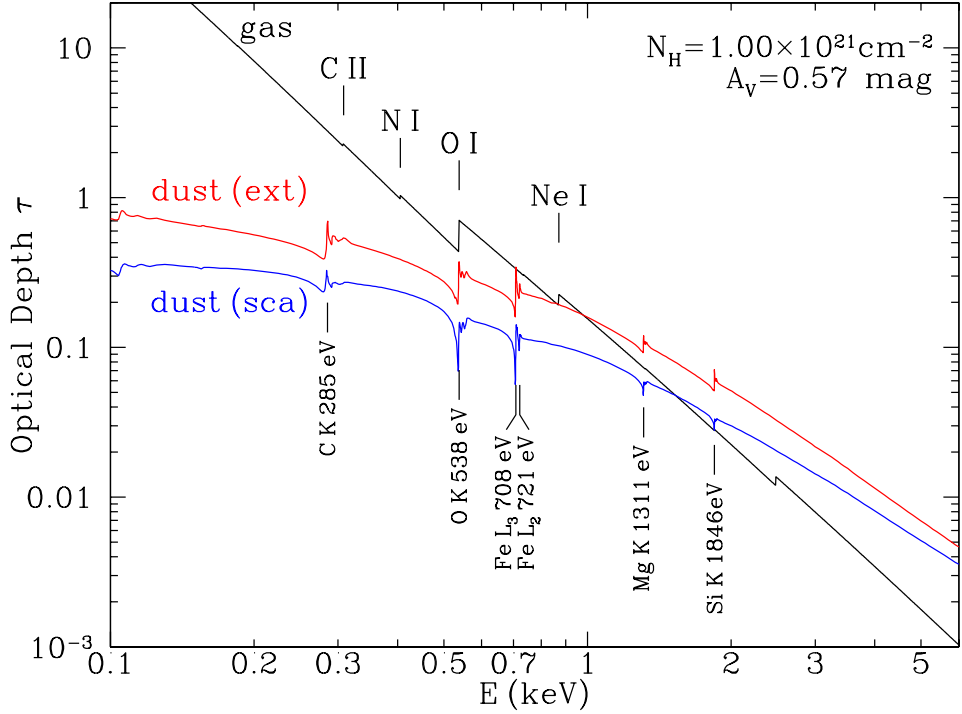}
\end{center}
\caption{
	\label{fig:ext}
	\footnotesize
	Optical depths due to photoelectric absorption by interstellar gas and
	extinction by dust, for a sightline with
	$N_{\rm H}^{\rm ISM}=1.0\times10^{21}\cm^{-2}$.
	Also shown is the contribution to dust extinction due to
	scattering.
	Photoelectric absorption edges are indicated for
	C~II, N~I, O~I, and Ne~I.
	}
\end{figure}

We will see below that the observed X-ray halo appears to be consistent
with models
with $N_{\rm H}^{\rm ISM}\approx 1.0\times10^{21}\cm^{-2}$
between us and the nova.
The absorption optical depth of the interstellar gas
is shown in Fig.\ \ref{fig:ext} as a function of energy, for
$N_{\rm H}^{\rm ISM} = 1.0\times 10^{21}\cm^{-2}$. 
Photoelectric absorption due to H, He, C, N, O, Ne, Mg, Si, S, and Fe
is included, with cross sections calculated following Verner and
Yakovlev (1995) and Verner et al.\ (1996), using subroutine {\tt phfit2.f}
written by D.A. Verner (1996).

In the interstellar gas,
abundances relative to H are taken to be 100\% of solar for He~I, N~I, Ne~I, 
and S~II,
30\% of solar for C~II, 80\% of solar for O~I,
and 10\% of solar for Mg~II, Si~II, and Fe~II due to depletion into
dust grains.
Nova Cygni 1992 did not form dust in the ejecta 
(Woodward et al.\ 1997).
In the circumstellar material we assume no dust grains, and solar abundances
for He~I, C~II, N~I, O~I, Ne~I, Mg~II, Si~II, S~II, and Fe~II.
Solar abundances for He, N, Ne, S are from Grevesse \& Sauval (1998);
for Si and Fe from Asplund (2000); for C from Allende Prieto et al.\ (2002a);
for O from Allende Prieto et al.\ (2002b).

In addition to gas phase absorption, there is absorption and
scattering by interstellar dust grains.  This is calculated assuming
a mixture of carbonaceous grains and silicate grains, with the
``case A'' size distribution found by WD01 for
Milky Way dust with $R_V\equiv A_V/E(B-V)=3.1$, but with grain abundance
per H nucleon reduced by a factor 0.93, as recommended by Draine (2003a).
Absorption and scattering cross sections were calculated using
dielectric functions which include structure near the X-ray absorption
edges (Draine 2003b), as described in \S\ref{sec:scatter}.

Extinction and scattering optical depths
for this dust are shown in Fig. \ref{fig:ext} as functions
of photon energy,
for a sightline with
$N_{\rm H}^{\rm ISM}=1.0\times10^{21}\cm^{-2}$.
Below $\sim0.5\keV$ absorption by the gas dominates, since H and He are
both assumed to be neutral.
At $\sim0.8\keV$ the dust grains provide $\sim$50\% of the extinction, and
at energies above $1\keV$ the dust grains dominate the extinction.

The interstellar matter ($N_{\rm H}^{\rm ISM}$)
and the
circumstellar gas ($\Delta N_{\rm H}$) are taken to have
differing attenuation properties, since some of the interstellar matter
(but none of the circumstellar material)
is in the form of dust grains.
Our estimate for the total attenuation therefore depends on the fraction of the
total column contributed by the interstellar medium, and this in turn
depends on the assumed distance $D$.

\section{X-Ray Spectrum of Nova V1974 Cygni 1992\label{sec:spectrum}}

Scattering of X-rays by dust grains is a function of the X-ray energy,
so it is important to use a realistic source spectrum when modelling this
phenomenon.
Mathis et al.\ (1995) approximated the emission as a blackbody
with $kT=22\eV$, attenuated by foreground gas with
$N_{\rm H}=4.25\times10^{21}\cm^{-2}$ and solar abundances,
while Witt et al.\ (2001) approximated the spectrum by a delta function
with $E=400\eV$.

Balman, Krautter \& \"Ogelman (1998) have recently examined the
lightcurve and spectrum of Nova Cygni 1992, and find that it appears
to be the sum of two separate components: thermal emission from
the photosphere of the white dwarf, with effective temperature $\Twd$
and radius $\Rwd$ both varying in time, plus emission from
a cooling and expanding thermal plasma.
We adopt this two-component model for the point-source flux $F_\nu$:
\beq
F_\nu(t) =
\left[F_\nu^{\rm pl}(t) + F_\nu^{\rm wd}(t)\right]
\exp\left[-\tau_{\rm abs,gas}(\nu) - \tau_{\rm abs,dust}(\nu) 
-\tau_{\rm sca}(\nu)\right]
~~~.
\eeq

\subsection{Thermal Plasma}

The flux $F_\nu^{\rm pl}$ from the emitting plasma is
\beq
F_\nu^{\rm pl} = \frac{1}{4\pi D^2} \int n_e n_\rmH dV 
\frac{\Lambda_\nu(T_{\rm pl})}{n_e n_\rmH}
\eeq
where $\Lambda_\nu(T_{\rm pl})$, 
the power radiated per unit volume and unit frequency by 
an optically-thin thermal plasma 
with kinetic temperature $T_{\rm pl}$,
is calculated using the 
Raymond \& Smith (1977) 
code for a
solar abundance plasma in collisional ionization
equilibrium.

We take the plasma properties to be given by
\beq
T_{\rm pl} = 
\left\{
	\begin{array}{ll}
	10^8 \K 				& t < 63 \days \\
	10^8\K\left[1 - 0.9(t-63\days)/42\days\right]	& 63\days < t < 105\days \\
	10^7 \K					& t > 105\days
	\end{array}
\right.
\label{eq:Tplasma}
\eeq
\beq
\label{eq:EM}
\frac{1}{4\pi D^2} \int n_e n_\rmH dV = 1.44\times 10^{12} 
A_{\rm pl}
\cm^{-5}
\left\{
	\begin{array}{ll}
	(t/94\days)^2	& t \leq 94\days	\\
	(t/94\days)^{-2}	& t > 94\days
	\end{array}
\right.
\eeq
with a correction factor
\beq
A_{\rm pl}=
\exp\left[
0.0925\left(\frac{N_{\rm H}^{\rm ISM}}{10^{21}\cm^{-2}}-1.0\right)
\right]
\eeq
depending on the assumed value of $N_{\rm H}^{\rm ISM}$.
The adopted plasma temperature (eq. \ref{eq:Tplasma})
is consistent with the temperature of this component inferred by
Balman et al., and eq. (\ref{eq:EM}) reproduces the \PSPC\ count rates
at times $t \leq 147\days$

\subsection{Nova Photosphere}

Balman et al.\ found that the emission from the nova photosphere was
consistent with the spectrum calculated by MacDonald \& Vennes (1991)
for O-Ne enhanced white dwarf model atmospheres.
Following Balman et al., we approximate the emission from the nova photosphere
using the O-Ne enhanced white dwarf model atmosphere spectra of
MacDonald \& Vennes (1991).
We will assume that the light curve consists of three phases:
(1) at times $t < t_{p1}$, the atmosphere is contracting at constant 
luminosity, with the temperature rising;
(2) at times $t_{p1} < t < t_{p2}$, the atmospheric radius remains constant,
with the temperature nearly constant (the ``plateau'' phase);
(3) at $t > t_{p2}$ the nova has exhausted its fuel, and the photosphere
begins to cool rapidly at constant radius.

The beginning and end of the plateau phase
are not well determined by the \ROSAT\ observations.
We will assume that the plateau phase began at $t_{p1}=335\days$,
and ended at $t_{p2}=565\days$.
The highest observed count rate was at $t=434\days$, at which time we
assume an effective temperature $kT_{\rm wd}=50\eV$;
the \ROSAT\ point source
count rate (see \S\ref{sec:model_countrate}) on day 434 can be reproduced if
\beq
\pi\left(\frac{\Rwd}{D}\right)^2=
2.61\times10^{-25}	
A_{\rm wd}
\sr ~~~{\rm for}~t>t_{p1}
\label{eq:OmegaWD}
\eeq
with a correction factor
\beq
A_{\rm wd}
=\exp\left[
0.198\left(\frac{N_{\rm H}^{\rm ISM}}{10^{21}\cm^{-2}}
-1.0
\right)\right]
\eeq
which depends weakly on the adopted value of $N_{\rm H}^{\rm ISM}$.
With the assumption of $R_{\rm wd}=$ constant for $t \geq t_{p1}$, we determine
the effective temperature $kT_{\rm wd}=49.84\eV$ on day 511.
We assume $dT_{\rm wd}/dt$ to be constant
throughout the plateau phase, giving $kT_{p1}=50.16\eV$, and
$kT_{p2}=49.72\eV$.
The peak luminosity is reached at $t_{p1}$:
\beq
L_{\rm wd} = 
2.57\times10^{38}	
A_{\rm wd}\left(\frac{D}{2.0\kpc}\right)^2\erg\s^{-1}
~~~.
\eeq
At times $t<t_{p1}$ the nova radius is obtained by assuming constant
luminosity:
\beq
\pi\left(\frac{\Rwd}{D}\right)^2 = 2.61\times10^{-25}
\left(\frac{kT_{p1}}{k\Twd} \right)^4 ~~~{\rm for}~t<t_{p1} ~~~.
\eeq
For $t =258$, 261, 612, 624, 635, and 650 days we determine $T_{\rm wd}$
by requiring that we reproduce the observed $\dot{N}_{ps}$, for
constant luminosity at $t < t_{p1}$, and constant radius for
$t > t_{p1}$.
For $t\leq 258\days$, we assume $T\propto t$.

Tabulated model atmosphere spectra have been obtained from MacDonald (2002)
for temperatures $T_{\rm wd}=1,2,3,3.5,4,5,6
\times10^5\K$; we
estimate the spectrum at intermediate temperatures by interpolation.

\subsection{Model Count Rate
	    \label{sec:model_countrate}}

Our aim is to reproduce the \ROSAT\ counts as a function of halo angle.
We must first ensure that our nova model reproduces the observed
count rates for the point-source component.
For the nova model and interstellar medium parameters described above, we
calculate the rate of 0.1 -- 2.4 keV photon detections by the \ROSAT\ \PSPC.
We use the \ROSAT\ effective area vs. energy from Snowden et al.\ (1994).

\begin{figure}[h]
\begin{center}
\epsfig{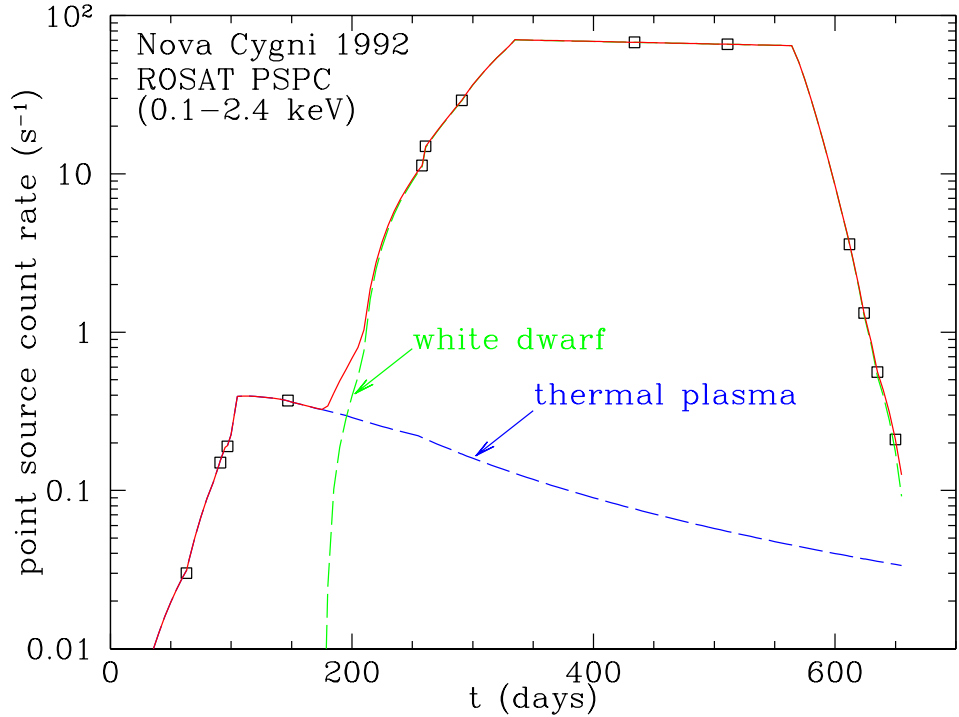}
\end{center}
\caption{
	\label{fig:lightcurve}
	\footnotesize
	Light curve for point source component of Nova Cygni 1992.
	Solid line: count rate for two component model,
	including attenuation by gas and dust.
	Broken curves: contributions from thermal plasma
	and white dwarf photosphere.
	Squares: measured count rates from Krautter et al.\ (1996)
	for $t\leq 147\days$, and from Table \ref{tab:count_rates}
	for $t\geq 258\days$.
	}
\end{figure}

In Figure \ref{fig:lightcurve} we show the \ROSAT\ point-source 
count rate calculated for
our model for the nova and intervening extinction, together with
measured count rates.

\begin{figure}[h]
\begin{center}
\epsfig{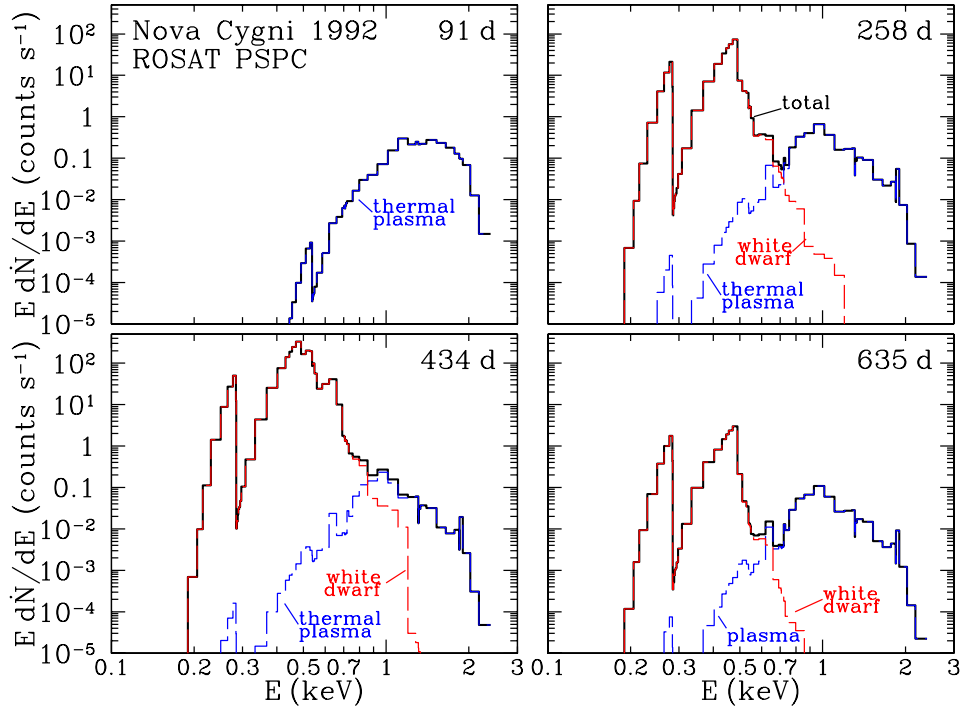}
\end{center}
\caption{
	\label{fig:ptsrc_spectra}
	\footnotesize
	\ROSAT\ \PSPC\ count rates as function of photon energy
	for two-component model for Nova Cygni 1992, including
	extinction by gas and dust, for
	$N_{\rm H}^{\rm ISM}=1.15\times10^{21}\cm^{-2}$.
	}
\end{figure}

In Figure \ref{fig:ptsrc_spectra} we show the calculated energy 
spectrum of {\it detected} 
photons at 4 different times. 
We note that the energy spectrum varies
considerably over the evolution of the nova.
At $t=91\days$ the spectrum is quite hard, being dominated by
the thermal emission from hot plasma with $kT_{\rm pl}\approx 3.4\keV$.
At this time the white dwarf photosphere is relatively cool, and the
radiation from it is absorbed by intervening H and He.
As the white dwarf photosphere contracts and becomes hotter, its 0.2--0.7~keV
emission
comes to dominate the count rate (see, e.g., the spectra for $t=258$, 434, and
$635\days$ in Figure \ref{fig:ptsrc_spectra}).

\begin{figure}[h]
\begin{center}
\epsfig{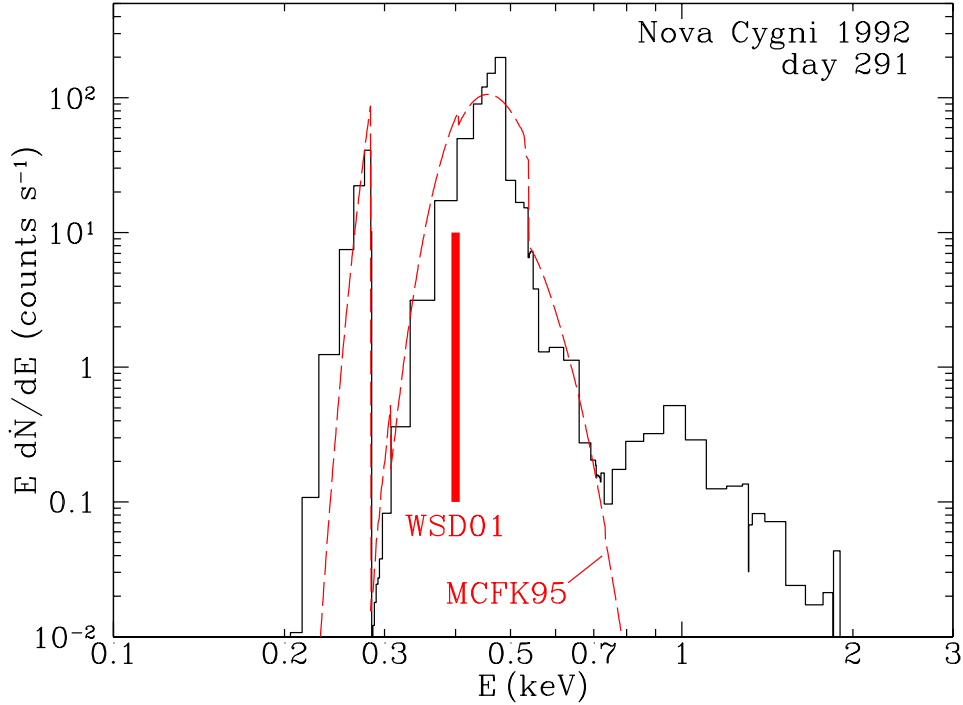}
\end{center}
\caption{
	\label{fig:compare_spec}
	\footnotesize
	\ROSAT\ \PSPC\
	count rate spectrum as function of photon energy
	for day 291.  Present model (solid histogram) is compared
	with count rate spectrum adopted by Mathis et al.\ (1995)
	(MCFK95) and the representative photon energy
	used by Witt et al.\ (2001) (WSD01).
	The drop in count rate at 284 eV 
	is due to absorption by C in the \ROSAT\ window.
	}
\end{figure}

In Fig.\ \ref{fig:compare_spec} we display the energy spectrum of
detected photons at $t=291\days$ after optical maximum, together
with the spectra used by Mathis et al.\ (1995) and Witt et al.\ (2001).

\section{X-Ray Scattering\label{sec:scatter}}

The theoretical framework for calculation of scattered halos 
around X-ray sources has been discussed by
Mauche \& Gorenstein (1986),
Mitsuda et al.\ (1990),
Mathis \& Lee (1991),
and Predehl \& Klose (1996).
The treatment which we present here is fully general, subject to only
the following approximations:

\begin{enumerate}
\item Polarization is neglected.  This is an excellent approximation
	for X-rays since scattering is only appreciable for small scattering 
	angles ($\ltsim1^\circ$) for which polarization effects are
	negligible.
\item Dust grains are assumed to be randomly-oriented.
	If dust grains are nonspherical and preferentially aligned,
	the dust scattering halo would not be azimuthally symmetric.
	While it may be possible to detect this effect in future
	observational studies, we neglect it in the present treatment.
	Generalization to include this effect would be straightforward
	but cumbersome, and not merited at this time.

\item We assume that the extinction (scattering by dust plus
	absorption by gas and dust) along each path contributing to the
	scattered intensity is the same as the extinction along the direct
	line from observer to source.  Since the scattering halos are small
	($\sim$90\% of the scattered photons within $\sim0.5^\circ$) this is
	quite a good approximation.  Towards Nova Cygni 1992 there are
	variations in $N_{\rm H}^{\rm ISM}(\infty)$ of $\sim 20\%$ on
	$0.5^\circ$ scales, particularly in a direction perpendicular to the
	Galactic plane (Hartmann \& Burton 1997). 
	In the optically-thin limit (i.e. for $h\nu\gtsim 1\keV$, see Fig.\ \ref{fig:ext}),
	such a density gradient has no effect on the azimuthally-averaged 
	halo intensity, relative to uniform extinction with a
	column equal to the average over the halo scale in the inhomogeneous case.
	When the optical depth is $\sim 1$ (i.e. for $h\nu\sim 500\eV$), 
	the differential absorption
	makes a small ($\sim 1\%$) effect on the azimuthally-averaged profiles 
	compared to the uniform case.

\item It is assumed that photons scattered by more than $90^\circ$
	may be neglected: we only consider scattered photons travelling
	away from the source (the ``outward-only approximation'').
\item The dust is
	assumed to be distributed spherically-symmetrically
	around the source.
	For example, there could be a spherical cavity surrounding
	the source.\footnote{%
		In our calculations we assume a small cavity around the source
		to avoid numerical difficulties with divergence in the
		integrand of eq.\ (\ref{eq:Itilde_1}) for $\theta\rightarrow0$
		and $y\rightarrow 1$.
		}
	Outside of this cavity,
	when we use a prescription 
	(e.g., exponential density law) for
	the dust density, we take this to be the density along the
	sightline to the nova; the assumed spherical symmetry
	provides densities away from the sightline.
	Within the small halo angles $\ltsim 1^\circ$ of interest,
	the resulting densities are close to what would have been
	obtained for the plane-parallel density structure which is
	the basis for our original density estimate.
	The assumed spherical symmetry implies azimuthal symmetry
	around the sightline to the source.
	Predehl \& Schmitt (1995) found that X-ray halos in \ROSAT\ observations
	did not show any detectable azimuthal asymmetries.
	
\end{enumerate}
Our formulae are presented without the small-angle
approximations which are frequently found in discussions of X-ray
scattering by dust.

Consider a point source of specific luminosity $L_\nu$ 
at a distance $D$ from the observer, and let $r$ be the distance
along the line from the observer to the source.
We assume the the dust size distribution and composition are the same
everywhere, but the dust spatial density $\rho$ may vary along
the line of sight.

Let
$\tau_{\rm abs}$ and $\tau_{\rm sca}$ be the total optical depths for scattering 
and extinction from source to observer.
At a distance $r=xD$ along the line from observer to the point source,
and at ``retarded'' time $t$
(time measured relative to the
arrival of a fiducial pulse emitted by the source), 
the intensity of $n-$times scattered photons arriving from
an angle $\theta$ is given by 
\beq
\label{eq:I_nu}
I_{n,\nu}(r,t,\theta)=
\frac{L_\nu(t)}{4\pi D^2}~e^{-\tau_{\rm abs}}
\frac{e^{-\tau_{\rm sca}}~\tau_{\rm sca}^n}{n!}
 ~\tilde{I}_n(x=r/D,t,\theta)
~~~.
\eeq
Eq.\ (\ref{eq:I_nu}) serves to define $\tilde{I}_n(x,t,\theta)$, which
depends on the dust distribution along the line-of-sight, but
not on the quantity of dust present.
The normalization is such that in the approximation of small
scattering angles, for a steady source one has
$\int \tilde{I}_n(0,t,\theta)2\pi\sin\theta d\theta=1$.

\begin{figure}[t]
\begin{center}
\epsfig{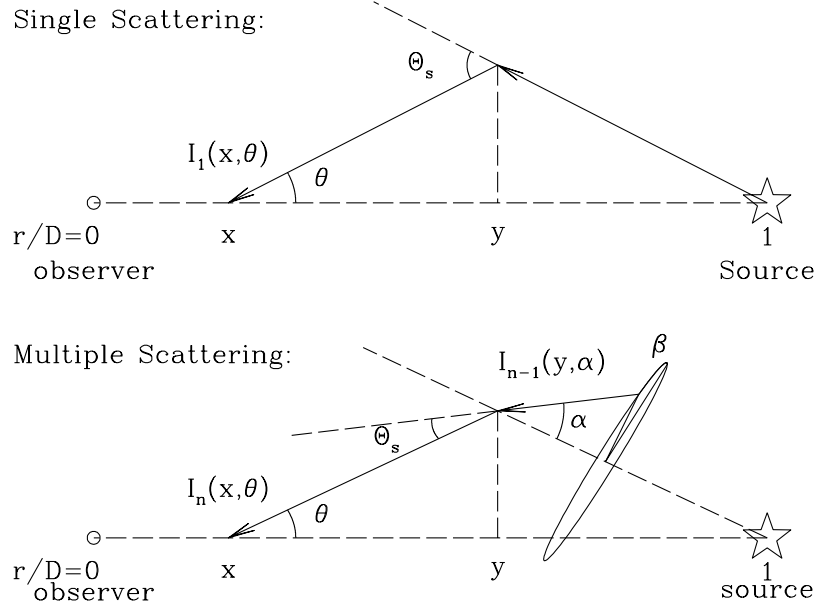}
\end{center}
\vspace*{-1.em}
\caption{
	\label{fig:geom}
	\footnotesize
	Geometry [upper panel] 
	for calculation of normalized single scattering intensity 
	$I_1(x,\theta)$
	and [lower panel] for calculation of multiple scattering intensity 
	$I_n(x,\theta)$
	given $I_{n-1}(y,\alpha)$ for $y>x$.
	The observer is at $x=0$.
	}
\end{figure}

\subsection{Single Scattering}

For single-scattering (see Fig.\ \ref{fig:geom})
the dimensionless intensity function $\tilde{I}_1$ is given by
\beq
\label{eq:Itilde_1}
\tilde{I}_1(x,t,\theta) = \frac{1}{\cos\theta}
\int_x^1 dy \frac{\tilde{\rho}(y^\prime)\tilde{\sigma}(\thetascat)}
		{(1-y)^2+(y-x)^2\tan^2\theta}
\frac{L_\nu(t-\delta t_1(x,y,\theta))}{L_\nu(t)}
\eeq
where
\beq
y^\prime \equiv 1 - \sqrt{(1-y)^2 + y^2\tan^2\theta}
\eeq
and the dimensionless density is
\beq
\tilde{\rho}(x)\equiv \frac{\rho(r=xD)}{D^{-1}\int_0^D\rho(r)dr}
~~~.
\eeq
The dimensionless differential scattering cross section is
\beq
\tilde{\sigma}(\thetascat)\equiv\frac{1}{\sigma_{sca}}\frac{d\sigma}{d\Omega}
~~~,\eeq
where $d\sigma/d\Omega$ is the differential scattering cross section
for scattering angle 
\beq
\label{eq:scatt_angle}
\thetascat(x,y,\theta) = \psi_0(x,y,\theta) 
\eeq
where
\beq
\label{eq:scatt_angle2}
\psi_0(x,y,\theta)\equiv 
\theta + \arctan\left[\frac{(y-x)\tan\theta}{1-y}\right],
\eeq
and the time delay
\begin{eqnarray}
\delta t_1 (x,y,\theta) &=& 
\frac{D}{c}
\left\{
\frac{y-x}{\cos\theta} 
+(1-y)\left[1+\left(\frac{y-x}{1-y}\right)^2\tan^2\theta\right]^{1/2}
-(1-x)\right\}
\\
&\approx& \frac{D}{c}\frac{(1-x)(y-x)}{(1-y)}\frac{\theta^2}{2}
= 28.0\days\left(\frac{D}{2.0\kpc}\right)\frac{(1-x)(y-x)}{(1-y)}
\left(\frac{\theta}{1000\arcsec}\right)^2.
\end{eqnarray}

For a steady source,
\beq
\tilde{I}_1 \rightarrow \tilde{I}_1^{\rm s}(x,\theta)
\equiv 
\frac{1}{\cos\theta}
\int_x^1 dy \frac{\tilde{\rho}(y^\prime)\tilde{\sigma}(\thetascat)}
		{(1-y)^2+(y-x)^2\tan^2\theta}
~~~.
\eeq

\subsection{Multiple Scattering}

For a steady source, the intensity of multiply-scattered photons 
is obtained from the recursion formula 
(Mathis \& Lee 1991; Predehl \& Klose 1996)
\beq
\tilde{I}_{n}^{\rm s}(x,\theta) = 
\frac{n}{\cos\theta}
\int_x^1 dy ~\tilde{\rho}(y^\prime) \int_0^{\alphamax}d\alpha\sin\alpha~
\tilde{I}_{n-1}^{\rm s}(y,\alpha)\int_0^{2\pi}d\beta~\tilde{\sigma}(\thetascat)
~~~(n \ge 2), \eeq
where the scattering angle
\beq
\thetascat(x,y,\theta,\alpha,\beta) = 
\arccos\left(\cos\alpha\cos\psi_0 + \sin\alpha\cos\beta\sin\psi_0\right)
~~~.
\eeq
The upper limit $\pi/2$ on the integral over $\alpha$ corresponds to
the ``outward-only'' approximation mentioned above.
For a nonsteady source, it is possible to
allow exactly for the time delays on different light paths, but the
formalism becomes cumbersome and the computations burdensome.
In the present application multiple scattering is a relatively minor
effect, and time delay is a relatively minor correction, 
so for $n\ge2$ we make only an approximate correction for
time delay by assuming that
the first $n-1$ scatterings occur
exactly midway between the source and the
last scattering:
\beq
\label{eq:Itilde_n}
\tilde{I}_{n}(x,t,\theta) \approx
\frac{n}{\cos\theta}
\int_x^1 dy ~\tilde{\rho}(y^\prime) \int_0^{\alphamax}d\alpha\sin\alpha~
\tilde{I}_{n-1}^{\rm s}(y,\alpha)
\left[\frac{L_\nu(t-\delta t_2)}{L_\nu(t)}\right]
\int_0^{2\pi}d\beta~\tilde{\sigma}(\thetascat)
~~~,
\eeq
\begin{eqnarray}
\delta t_2 (x,\theta,\alpha) &=&
\frac{D}{c}
\left\{\frac{x}{\cos\theta} + 
\frac{\left[(1-x)^2+x^2\tan^2\theta\right]^{1/2}}{\cos\alpha} -1 \right\}
\\
&\approx& \frac{D}{2c}\left[\frac{x}{(1-x)}\theta^2+(1-x)\alpha^2\right]
~~~.
\end{eqnarray}

To calculate the intensity $I_n(0,t,\theta)$ of $n$-times scattered photons
at $x=0$, we precalculate and tabulate 
$\tilde{I}_{n-1}^s(y,\theta)$ at selected
values of $y$ and $\theta$, and then obtain $\tilde{I}_{n-1}^s$ by
interpolation when evaluating eq.\ (\ref{eq:Itilde_n}) for
$\tilde I_n(0,x,t)$.

\begin{figure}[h]
\begin{center}
\epsfig{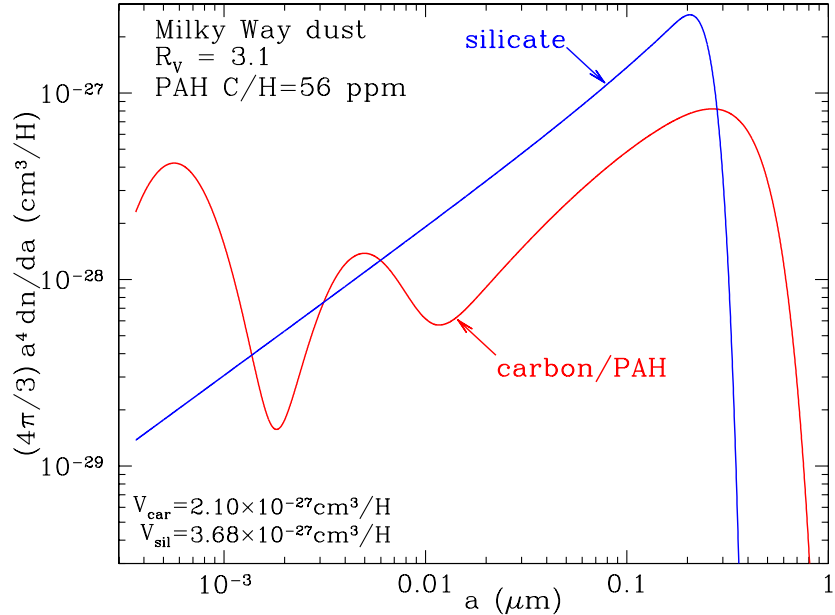}
\end{center}
\vspace*{-2em}
\caption{\label{fig:dnda}
	\footnotesize
	WD01
	size distributions for carbonaceous grains (including PAHs)
	and amorphous silicate grains for Milky Way dust with $R_V=3.1$ and
	C/H=60ppm in PAHs.
	Total volume of carbonaceous grains and silicate grains per H is 
	$2.26\times10^{-27}$ and $3.96\times10^{-27}\cm^3$,
	respectively.
	}
\end{figure}
\begin{figure}[h]
\begin{center}
\epsfig{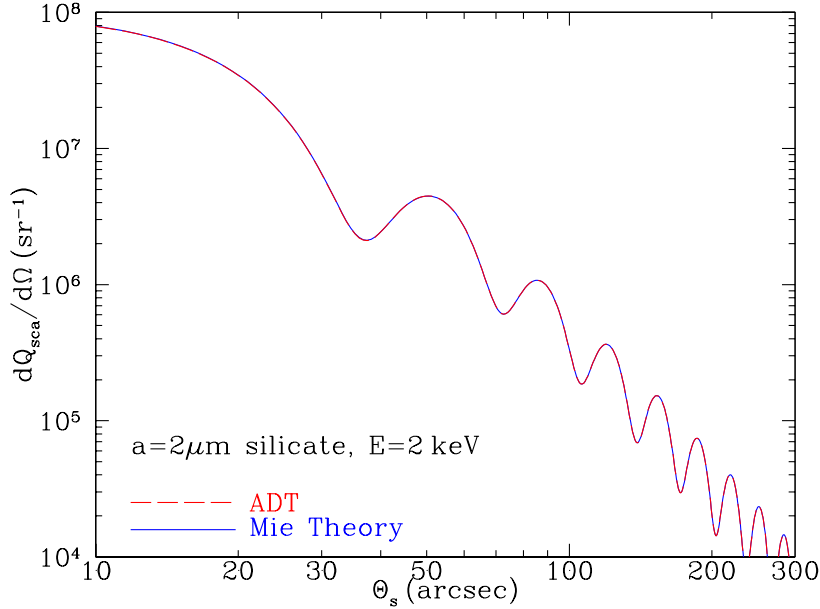}
\end{center}
\vspace*{-2.em}
\caption{
	\label{fig:adt}
	\footnotesize
	Differential scattering cross section calculated using Mie theory
	and anomalous diffraction theory for $a=2\micron$ silicate grain
	at 2 keV, for which $x=2.03\times10^4$.  The two curves are
	indistinguishable.
	}
\end{figure}

\subsection{Dust Model and Scattering Cross Sections}

For the line of sight to Nova Cygni 1992, we assume that the dust is
``average'' diffuse cloud dust, and we
adopt the dust model developed by 
WD01 for $R_V\equiv A_V/E(B-V)\approx3.1$ and C/H~=~60~ppm in polycyclic 
aromatic hydrocarbons
(PAHs), except that, following Draine (2003a), 
the abundances of all grain components (relative to H) are reduced,
by a factor 0.93; with this adjustment, the grain
model reproduces the
estimated extinction per H nucleon.
The model consists of a mixture of carbonaceous grains and
amorphous silicate grains.
As discussed by Li \& Draine (2001), 
the carbonaceous grains have the properties of 
PAH molecules when they contain $\ltsim 10^4$ C atoms, and the properties
of graphite particles when they contain $\gtsim 10^5$ C atoms.
By altering the size distributions of the carbonaceous and silicate particles,
the dust model appears to be able to reproduce observed extinction curves
in various Galactic regions, and in the Large and Small Magellanic
Clouds.
This dust model, when illuminated by starlight, produces infrared emission
consistent with the observed emission spectrum of the interstellar medium
(Li \& Draine 2001, 2002).
The size distribution for this dust model is shown in Fig.\ \ref{fig:dnda}.
For assumed densities of $2.2\g\cm^{-3}$ for the carbonaceous grains
and $3.8\g\cm^{-3}$ for the MgFeSiO$_4$ silicate grains, this corresponds to a 
dust-to-gas mass ratio of 0.008

The scattering properties of a spherical target of radius $a$
are determined by the
complex refractive index $m(\lambda)$ and the dimensionless
size parameter $x\equiv 2\pi a/\lambda$.
Many early papers on X-ray scattering by dust employed the ``Rayleigh-Gans''
approximation to calculate the differential scattering cross sections.
Smith \& Dwek (1998) showed, however, that the 
Rayleigh-Gans validity criterion $|m-1|x\ll 1$
is not satisfied for interstellar
grains at energies $\ltsim$1~keV.
In the present work we assume spherical, homogeneous grains and
use exact Mie scattering theory 
(see, e.g. Bohren \& Huffman 1984) to calculate differential scattering
cross sections $d\sigma/d\Omega$,
employing a computer code based upon the subroutine MIEV0
written by Wiscombe (1980, 1996), modified to use IEEE 64 bit arithmetic.
Wiscombe's code is accurate for ``size parameters''
$x = 7800 (a/\micron)(\keV/h\nu) < 2\times10^4$.
For $x>2\times10^4$ we calculate $d\sigma/d\Omega$ 
using ``anomalous diffraction theory'' (van de Hulst 1957), since
the validity criteria $x\gg 1$ and $|m-1|\ll 1$ are satisfied.
The accuracy of the scattering calculations is illustrated in 
Fig.\ \ref{fig:adt}, where we show the differential scattering cross
section calculated for a case with $x=2\times10^4$ -- the Mie theory
calculation and the anomalous diffraction theory calculation are
indistinguishable.
Since the two calculations are entirely different in approach, the
agreement confirms that both are accurate.

\begin{figure}[h]
\begin{center}
\epsfig{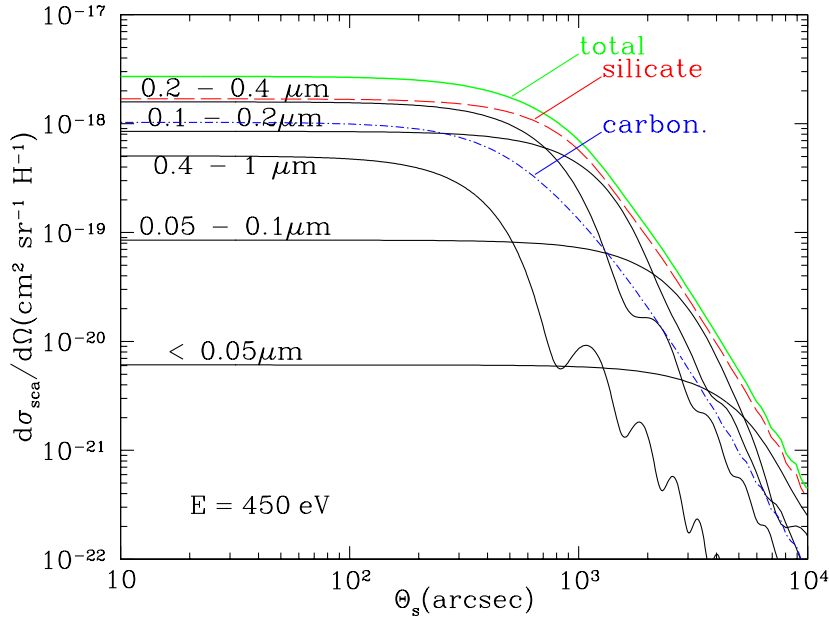}
\end{center}
\vspace*{-2.em}
\caption{
	\label{fig:scat}
	\footnotesize
	Heavy solid curve: 
	differential scattering cross section per H at
	$h\nu=450\eV$ for the WD01 model for Milky Way dust
	with $R_V=3.1$ and C/H=60 ppm in PAHs.
	Also shown are separate 
	contributions of silicate and carbonaceous grains,
	and contributions from different size ranges.
	}
\end{figure}

For the amorphous silicate grains and the carbonaceous grains we employed
the dielectric functions estimated by Draine (2003b)
for olivine MgFeSiO$_4$ and graphite, respectively.
Figure \ref{fig:scat} shows the
differential scattering cross section per
H nucleon for the dust mixture of WD01 for
Milky Way dust with $R_V=3.1$ (with grain abundances reduced by a factor
0.93).
For this mixture we see that the scattering 
for scattering angles $\Theta_s<1000$\arcsec\  
is dominated by grains with
radii in the $0.1-0.4\micron$ range -- larger grains contribute only
$\sim$20\% of the total scattering cross section for 
$\Theta_s \ltsim 200$\arcsec, and $\ltsim 1$\% for $\Theta_s\gtsim 500$\arcsec.
We also see that silicate grains provide $\sim$60\% of the scattering
for $\Theta_s\ltsim300$\arcsec, increasing to $\sim$80\% at 
$\Theta_s\gtsim1000$\arcsec.

\begin{figure}[h]
\begin{center}
\epsfig{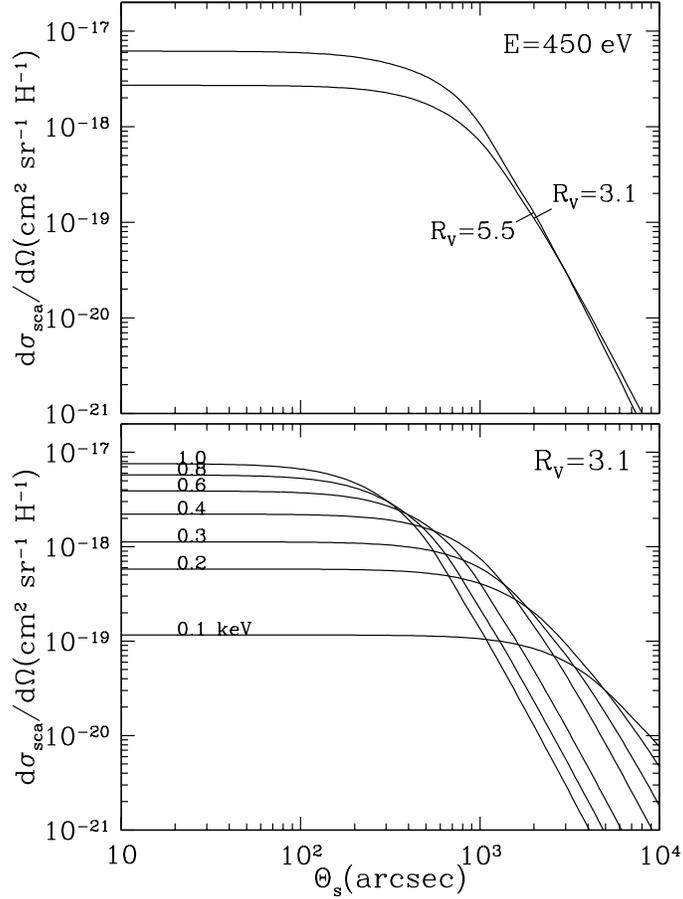}
\end{center}
\caption{
	\label{fig:8en_RV}
	\footnotesize
	Upper panel: differential scattering cross section at $E=450\eV$
	for WD01 model for Milky Way dust with $R_V=3.1$ (diffuse cloud
	average) and $R_V=5.5$ (dense cloud).
	Lower panel: differential scattering cross sections at 8 energies
	for WD01 model for Milky Way dust with $R_V=3.1$.
	}
\end{figure}

It is well-known that the extinction curve varies from one region to another,
which is presumed to be due to changes in the size distribution of the dust.
In addition to the size distribution which reproduces the standard
Milky Way diffuse cloud extinction curve with $R_V=3.1$,
WD01 have constructed a size distribution consistent with an extinction
curve with $R_V=5.5$.
Such flat extinction curves are found in dense
regions, and require a shift of the dust size distribution toward larger
sizes.
The X-ray scattering properties for dust with $R_V=3.1$ and 5.5 are compared
in Fig.\ \ref{fig:8en_RV} -- we see that $R_V=5.5$ dust is
slightly more forward-scattering than $R_V=3.1$ dust.
The cross section for a scattering angle of $\ltsim 300\arcsec$ is
larger by about a factor 1.4.

The scattering also depends on the X-ray energy.
In Fig.\ \ref{fig:8en_RV} we show the differential scattering properties
of $R_V=3.1$ dust at energies ranging from 0.1 to 1 keV.
As the energy increases, the dust becomes more forward-scattering.


\begin{figure}[h]
\begin{center}
\epsfig{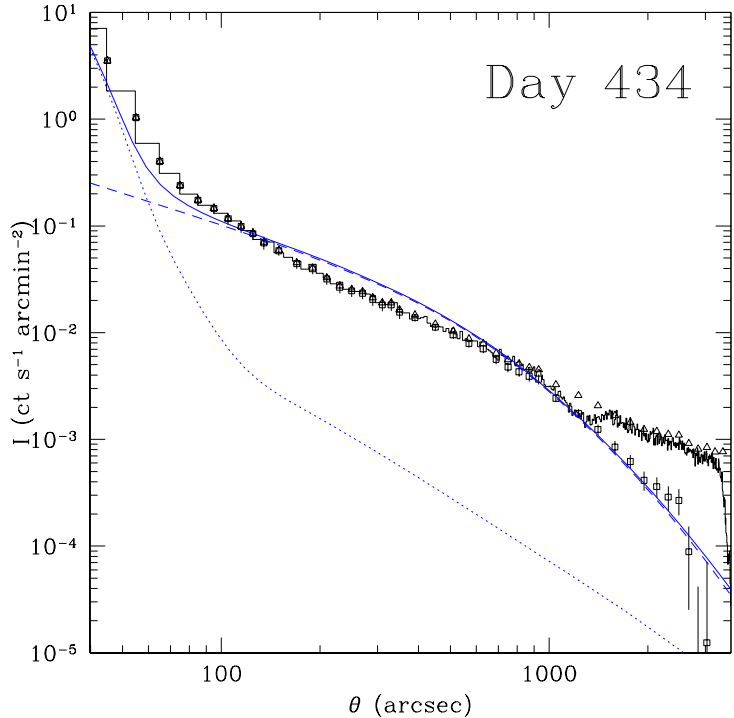}
\end{center}
\vspace*{-2.5em}
\caption{
	\label{fig:day434}
	\footnotesize
	X-ray intensity profile from Nova Cygni 1992, 434 days after
	optical maximum.
	Solid histogram: raw \ROSAT\ \PSPC\ data in 10\arcsec bins. 
	Triangles: intensities
	after processing by ESAS software (see text).
	The background is determined by averaging the 
	intensity in the annulus from 2800-3200\arcsec. 
	Squares: nova intensity after background subtraction. 
	The error bar around
	each point shows the $3-\sigma$ statistical error but does not include
	any uncertainty due to background subtraction.
	Solid curve: model
	intensity profile [psf (dotted line) 
	plus dust-scattered halo (dashed line)] for
	nova at $D=2.1\kpc$.
	The model is in generally good agreement over the
	full 40--3000\arcsec range where the scattered halo can be measured,
	but overpredicts the scattered intensity by $\sim$40\% 
	at $\theta\approx250$\arcsec,
	and underpredicts the observed intensity at $\theta < 100\arcsec$.
	}
\end{figure}

\begin{figure}[h]
\begin{center}
\epsfig{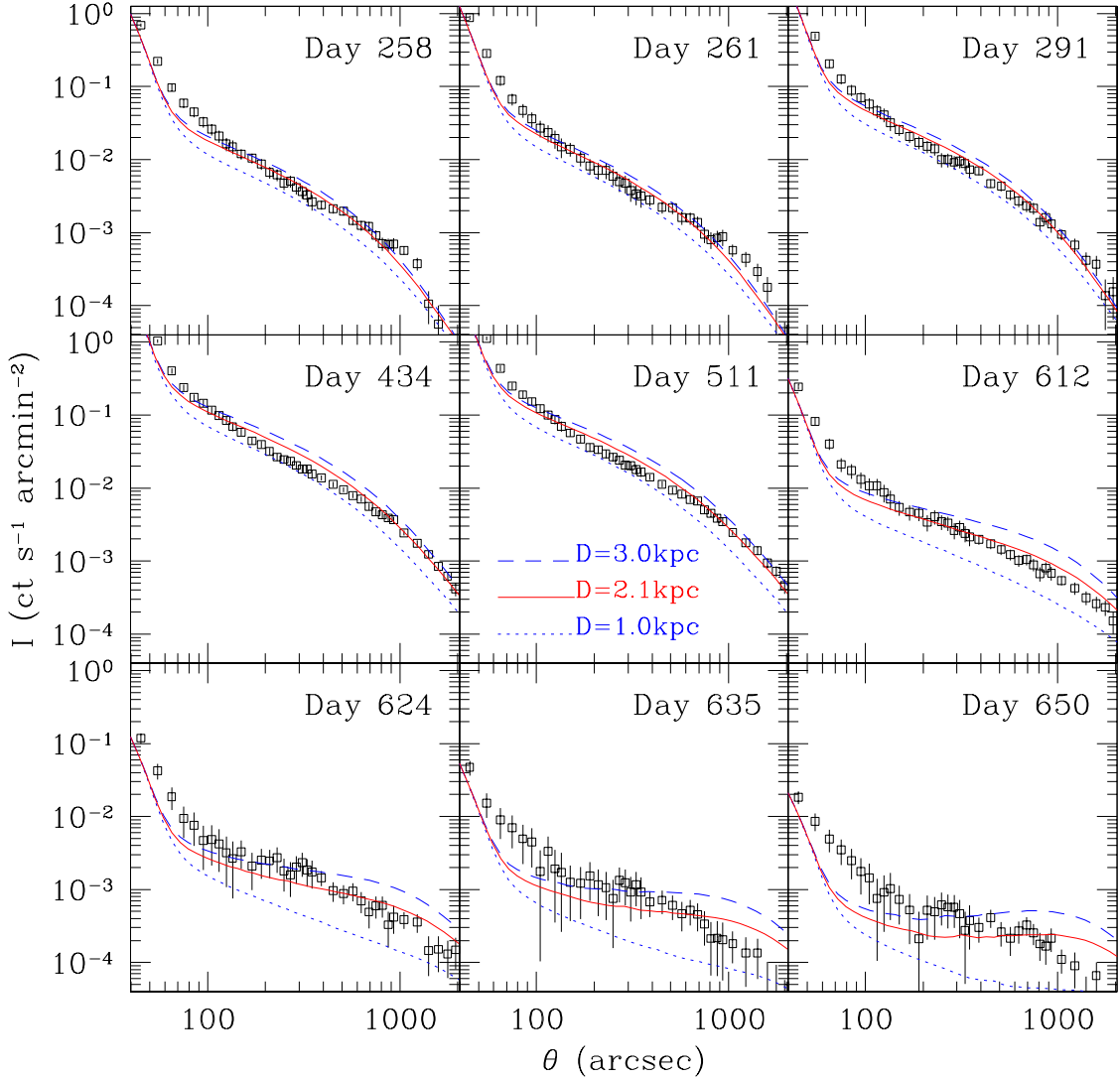}
\end{center}
\vspace*{-2.5em}
\caption{
	\label{fig:halo_mult_D}
	\footnotesize
	Multi-epoch X-ray intensity profiles from Nova Cygni 1992.
	Squares: observed intensity profile; the error bar shows the
	$3-\sigma$ statistical uncertainty.
	Solid lines: models consisting of psf plus
	dust-scattered halo for exponential density law with
	$z_{1/2}=300\pc$ and nova at an assumed distance of $D = 2.1\kpc$.
	This distance, corresponding to $N_{\rm H}^{\rm ISM}=1.15
	\times 10^{21}\cm^{-2}$, or $E(B-V)=0.20$~mag, gives the best 
	overall match to the data.
	Dotted lines: the same models but for 
	$D = 1.0$ (lower line) and $3.0\kpc$ (upper line)
	(corresponding to $N_{\rm H}^{\rm ISM}=0.64$ and $1.46\times
	10^{21}\cm^{-2}$, or $E(B-V)=0.11$ and $0.25$~mag.
	}
\end{figure}

\section{Results\label{sec:results}}

\subsection{Calculations}

Using the method described in \S\ref{sec:scatter}, the
dust-scattered halo was calculated at each of 9 epochs, and
the energy-dependent \ROSAT\ \PSPC\ psf of
Boese (2001) was used to model the 
contribution of the unscattered photons to the image.
We include the contributions to the scattered halo from singly and
doubly-scattered photons.
For the scattering optical depth 
$\tau_{\rm sca}< 0.3$ for $E>250\eV$ (see Fig.\ \ref{fig:ext}),
photons scattered 3 or more times contribute only a fraction
\beq
\frac{\frac{1}{3!}\tau_{\rm sca}^3+\frac{1}{4!}\tau_{\rm sca}^4+...}
{e^{\tau_{\rm sca}}-1} <.014
\eeq
of the total halo counts.

We use 71 energy bins extending from $0.1$ to $2.4\keV$, 
chosen so that absorption edge structure is well-defined.
The dust size distribution is treated using 50 size bins spanning the
range $.00035\micron < a < 1.0\micron$ (except for the ``big grain''
case discussed below, where we use 55 bins running from 
$.00035\micron < a < 2.0\micron$).
We calculate the scattering halo at 80 different halo angles $0 < \theta <
10^4$arcsec.  The dust scattering properties $\tilde{\sigma}(\Theta_s)$
are precalculated at
115 different scattering angles $\Theta_s$, with interpolation used
during evaluation of $\tilde{I}_1$ [eq.\ (\ref{eq:Itilde_1})]
and $\tilde{I}_2$ [eq.\ (\ref{eq:Itilde_n})].

For computation of the second-order scattering, $I_1^{\rm s}(x,\theta)$
is evaluated at 40 different locations $x$, with interpolation used
for evaluation of $\tilde{I}_2$.
We have verified the accuracy of the numerical quadratures by
doubling the number of spatial steps and confirming that the results
do not change significantly.

Our preferred exponential model, with
an assumed distance $D=2.1\kpc$,
is shown in Figure \ref{fig:day434},
including the separate contributions of the point source (broadened by
the psf) and the scattered halo.

\subsection{Comparison with Observational Data}

The radial distributions of \ROSAT\ \PSPC\ counts are shown in
Figure \ref{fig:halo_mult_D} for the nine epochs for which
high quality \ROSAT\ data are available (see \S\ref{sec:XRayObs}).

As discussed in \S\ref{sec:XRayObs}, our method of
evaluating the background forces the derived halos to artificially go
to zero at 3000\arcsec. 
Thus the observed halo intensities are only reliable for
intensities significantly 
greater than the true halo intensity at 3000\arcsec\ and
for this reason we restrict the useful comparisons of models with data
to the region inside $\sim 2000$\arcsec. An example of the raw,
processed, and background-subtracted data is shown in Figure
\ref{fig:day434} for day 434, which had the highest count rate
($67.8\ct\s^{-1}$) and number of counts ($\sim$200,000`) of unscattered
photons from the nova.

We
also show models calculated for several values of the nova
distance $D$, ranging from $D=1.0$ to $3.0\kpc$ (holding fixed
the assumed exponential
gas distribution, with $z_{1/2}=300\pc$).
As $D$ is increased,
the dust scattering optical depth $\tau_{\rm sca}$ to the nova increases,
leading to an increase in the intensity of the scattered X-ray halo.
Note that as $D$ is varied the observed point source 
flux is held constant by scaling the intrinsic nova luminosity.

To quantify the goodness-of-fit, for each of the epochs $k=1$--9
we divide the range
50--2040\arcsec\ into annuli, $j=1$--36,
with widths $\Delta\theta=10$\arcsec\ for 50--140\arcsec,
20\arcsec\ for 140--360\arcsec,
60\arcsec\ for 360--960\arcsec,
and 180\arcsec\ for 960--2040\arcsec
(the nova data are plotted in these intervals in Figures \ref{fig:day434} 
and \ref{fig:halo_mult_D}).
Let $I_{j,k}^{\rm obs}$ be the observed intensity
(counts s$^{-1}$ arcmin$^{-2}$) 
including the background
for annulus $j$ and epoch $k$,
let $I_{j,k}^{\rm bkg}$ be the estimated background intensity in
annulus $j$ at epoch
$k$, and let $I_{j,k}^{\rm mod}$ be the intensity
calculated for the model at epoch $k$.
We use a goodness-of-fit metric
\beq
\chi^2 = 
\sum_{k=1}^9 \sum_{j=1}^{36} 
\frac{\left[I_{j,k}^{\rm obs}-I_{j,k}^{\rm bkg}-I_{j,k}^{\rm mod}\right]^2}
{\sigma_{j,k}^2}
~~~,
\eeq
\beq
\label{eq:sigma2}
\sigma_{j,k}^2\equiv
	I_{j,k}^{\rm obs}/(\Omega_j\Delta t_k)
+ 0.01\left(I_{j,k}^{\rm bkg}\right)^2
+ 0.04\left(I_{j,k}^{\rm obs}-I_{j,k}^{\rm bkg}\right)^2
+ 0.04\left(I_{j,k}^{\rm mod}\right)^2
~~~.
\eeq
The first term in eq.\ (\ref{eq:sigma2}) 
is due to the statistical uncertainty in the
number of photons counted ($\Omega_j$ is the solid angle of annulus $j$, and
$\Delta t_k$ is the exposure time for epoch $k$).
The second term allows for an estimated $\pm10\%$
uncertainty in the estimated background level.
The third and fourth terms
are somewhat arbitrary, but are introduced to avoid overly weighting
the regions of the halo where the photon statistics may be good,
but where there may still be unknown systematic errors in the
observations, as well as systematic errors in the models due to inaccuracies
in the adopted nova light curve, nova spectrum, 
grain dielectric functions, etc.
Note, for example,
that even though the count rate appears to be systematically rising
between days 255 and 292 (Krautter et al.\ 1996) the measured
count rate on day 292 is 11\% smaller than the count rate measured on
day 291,
indicating either a variation in instrumental response or complexities
in the nova emission which are not allowed for in our model.
With $\sigma^2$ defined by eq.\ (\ref{eq:sigma2}),
a single data point cannot contribute more than 25 to $\chi^2$.
\begin{figure}[h]
\begin{center}
\epsfig{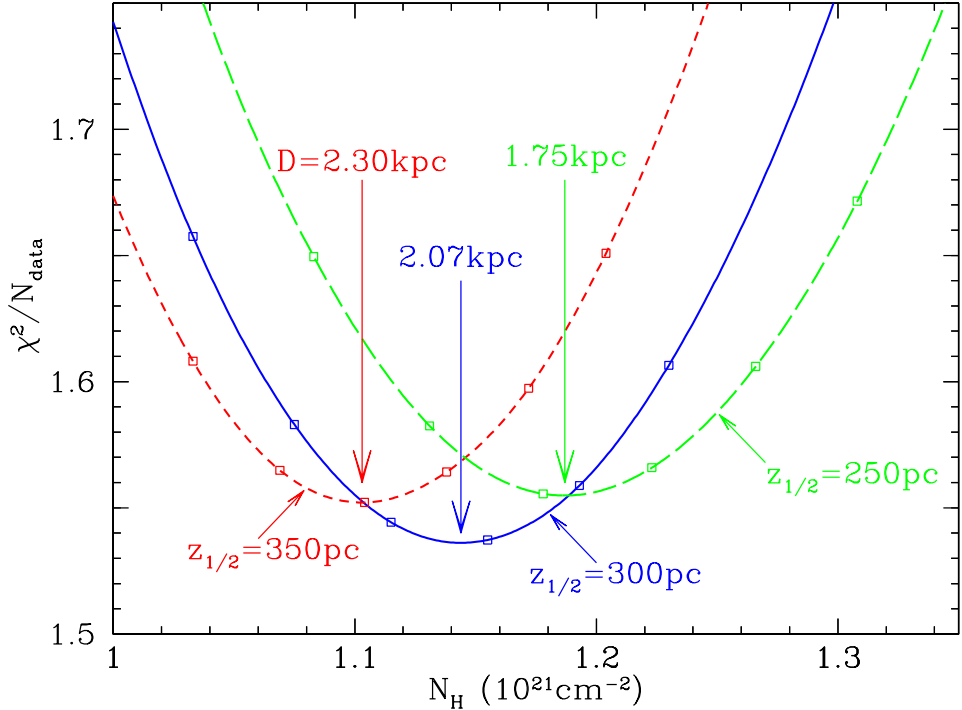}
\end{center}
\vspace*{-2.em}
\caption{
	\label{fig:chisq.NH}
	\footnotesize
	Figure-of-merit $\chi^2$ as a function of gas column density
	$N_{\rm H}^{\rm ISM}$.  
	All our models favor an interstellar column density
	$N_{\rm H}^{\rm ISM}\approx 1.1-1.2\times10^{21}\cm^{-2}$ 
	toward Nova Cygni 1992, and $E(B-V)\approx 0.19 - 0.21$~mag.
	}
\end{figure}

Fig.\ \ref{fig:chisq.NH} shows $\chi^2$ as a function of
$N_{\rm H}^{\rm ISM}$ for three exponential density laws.
For $z_{1/2}=300\pc$,
$\chi^2$ is minimized for
column density
$N_{\rm H}^{\rm ISM}\approx1.15\times10^{21}\cm^{-2}$,
corresponding to a distance $D\approx2100\pc$.
This model is shown in Figs. \ref{fig:day434} and
\ref{fig:halo_mult_D} by the solid line.
At shorter distances $N_{\rm H}^{\rm ISM}$ is smaller and
the model halo is systematically too weak, while
for larger distances $N_{\rm H}^{\rm ISM}$ is larger and
the model halo is sytematically too strong.
The best-fit model still shows systematic discrepancies with the observations
 -- it can be seen from Fig.\ \ref{fig:halo_mult_D}
that the model tends to overpredict the
halo intensity for 200--500\arcsec\ for $t\leq511\days$,
and underpredicts the halo intensity for 50--100\arcsec\ at all times 
-- but the model halo profiles are in generally good agreement
($\pm \lesssim 40\%$)
with the data at all epochs.

\begin{figure}[h]
\begin{center}
\epsfig{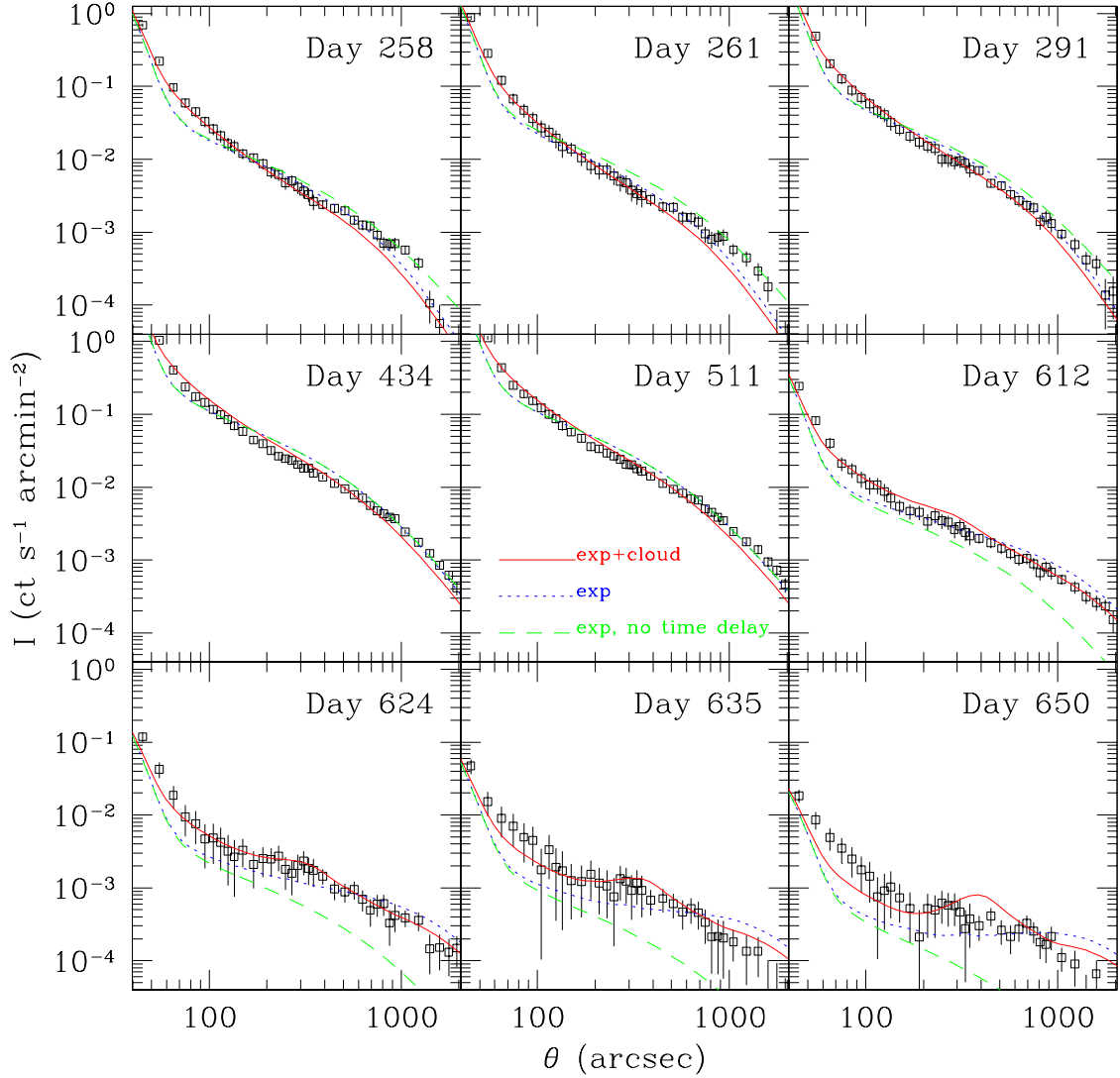}
\end{center}
\vspace*{-2.5em}
\caption{
	\label{fig:delays}
	\footnotesize
	Effect of time delays 
	and of adding a cloud close to the nova: 
	Same as Fig.\ \ref{fig:halo_mult_D} but showing $D=2.1$~kpc
	model (dotted line) with same model calculated neglecting
	time delay effects (dashed line).
	At early times ($t\leq 291\days$) the effect of time delay
	is to lower the intensity at large angles;
	at late times (lower panel) the effect of time delay is
	to increase the halo intensity at large angles.
	A ``cloud + exponential'' model, also for $D=2.1$~kpc, 
	with 70\% of dust in the exponential distribution
	and 30\% in a cloud at $y=r/D=0.95$ is shown by the solid line (see 
	also Figs.\ \ref{fig:clump} \& \ref{fig:biggrains} and \S\ref{sec:ddist})
	}
\end{figure}

Fig.\ \ref{fig:chisq.NH} shows that exponential
density laws with different values
of $z_{1/2}$ are nearly degenerate:
the best-fit distance in each case corresponds to the distance
at which the column density $N_{\rm H}(D)\approx 1.15\times10^{21}\cm^{-2}$,
or $D\approx 2100 (z_{1/2}/300\pc)\pc$.
For a steady source these solutions would in fact be perfectly
degenerate -- the difference between models with the same
$N_{\rm H}^{\rm ISM}$ but different $z_{1/2}$ is due to the
variability of the source and the different time delays for
models with different $D$.
The $\chi^2$ curves for the 3 different exponential distributions do
differ slightly, but
evidently the differences in time delay 
have a relatively small effect on the overall $\chi^2$.
This is because the effect of time delays on the halos is only
prominent at late times when the lightcurve is declining and at large
angles where the fluxes are close to the level of the background (see
Fig. \ref{fig:delays}). 

Fig.\ \ref{fig:delays} shows X-ray halos at each epoch for our standard model,
as well as the results for the same model but with time delays set to zero
(e.g., infinite speed of light).
Time delay effects are minimal at days 434 and 511 because the
light curve (Fig.\ \ref{fig:lightcurve})
is relatively constant for $\gtsim$50 days prior to
those two epochs; at earlier epochs ($t\leq 291\days$) 
the effect of time delay is
to reduce the intensity at large angles, and at $t \ge 612\days$
the effect of time delay is to increase the intensity at large
angles.

\subsection{Reddening}

Our modelling of the X-ray halos favors a column density
$N_{\rm H}^{\rm ISM}\approx 1.15\times10^{21}\cm^{-2}$ 
in order to
get the overall halo intensity correct; this column of gas and dust
corresponds to a reddening 
$E(B-V)\approx0.20$~mag.
How does this compare to other reddening estimates?

Barger et al.\ (1993) measured the
H$\alpha$/H$\beta$ intensity ratio at different times.
The ratio declined with time, presumably due to the nebula becoming
optically thin in the Balmer lines.
The last measurement reported 
by Barger et al.\ ($t=450\days$ after optical maximum),
was
$I({\rm H}\alpha)/I({\rm H}\beta)=3.47$; 
Mathis et al.\ (1995) make a
reasonable extrapolation to an asymptotic value $3.31$, from
which they estimate $E(B-V)=0.19$~mag.
If the nebula is optically thick in the Balmer lines, radiative transfer
effects can lead to an increase in the H$\alpha$/H$\beta$ ratio relative
to the ``case B recombination'' value which is assumed; this would cause
$E(B-V)$ to be overestimated.  Similarly, at high densities there can
be $n=2\rightarrow3$ collisional excitation, which would again increase
H$\alpha$ emission and lead to overestimation of $E(B-V)$.
Thus the H$\alpha$/H$\beta$ estimate of $E(B-V)$ should be regarded as
an upper bound.  Since the H$\alpha$/H$\beta$ estimate of $E(B-V)$ is
in agreement with our estimate, it appears that radiative trapping
effects do not appreciably affect the H$\alpha$/H$\beta$ ratio at
$t\gtsim 450\days$.

Austin et al.\ (1996) estimate $E(B-V)=0.40\pm0.07$ using the
He~II~1640/4686 line ratio, and $0.38\pm0.07$ using the
[Ne~IV]~1602/4724 line ratio.  In principle these line ratios should
be reliable reddening indicators.\footnote{%
	The [Ne~IV]
	line ratio should be essentially independent of nebular conditions
	(Draine \& Bahcall 1981).
	If H$\alpha$/H$\beta$ is not appreciably
	affected by radiative trapping effects, then the He~II~1640/4686 ratio
	would be expected to also be consistent 
	with case B recombination theory.
	}
However, the He~II and [Ne~IV] reddening estimates both rely on IUE
photometry in the 1600-1640\AA\ range, and we note that the
He~II~1640/4686 and [Ne~IV]~1602/4724 ratios measured by Austin et al.\
at 5 epochs are strongly correlated, which may indicate calibration
errors in the IUE spectrophotometry.

If the interstellar gas column on the sightline to the nova substantially
exceeds $1.15\times10^{21}\cm^{-2}$ and the true reddening significantly
exceeds $0.20$~mag,
the WD01 dust model will be disfavored:
if the column density of interstellar dust and gas is appreciably
increased, the predicted halo intensities
become too large at most
angles and most epochs, and the goodness-of-fit suffers
(see Fig.\ \ref{fig:chisq.NH}).
We conclude that
the WD01 dust model is incompatible
with $E(B-V)\gtsim 0.3$ in smoothly-distributed dust toward the nova.
If $E(B-V)$ is indeed as large as 0.36 
(the value recommended by Austin et al.)
then either 
an appreciable fraction of the reddening must be contributed by dust
which is located close to the nova
(e.g., dust associated with the nova, or interstellar 
dust within $\sim$50~pc of the nova) which
would contribute only to very small halo angles, and would be 
indistinguishable from the instrumental
point spread function)
or the WD01 model must be rejected.
However, it seems most likely that the reddening is close to 
$E(B-V)\approx 0.19$~mag, the value obtained from the H$\alpha$/H$\beta$
ratio.\footnote{%
	From the similar colors of nova V1974 Cygni 1992 and
	nova V382 Velorum 1999 (for which the H column has been measured
	to be $N_{\rm H}\approx 1.2\times10^{21}\cm^{-2}$)
	Shore et al.\ (2003) conclude that
	the reddening to nova V1974 Cygni 1992 is probably in the
	range 0.2-0.3.}

\subsection{Dust Distribution Along the Line-of-Sight\label{sec:ddist}}

\begin{figure}[h]
\begin{center}
\epsfig{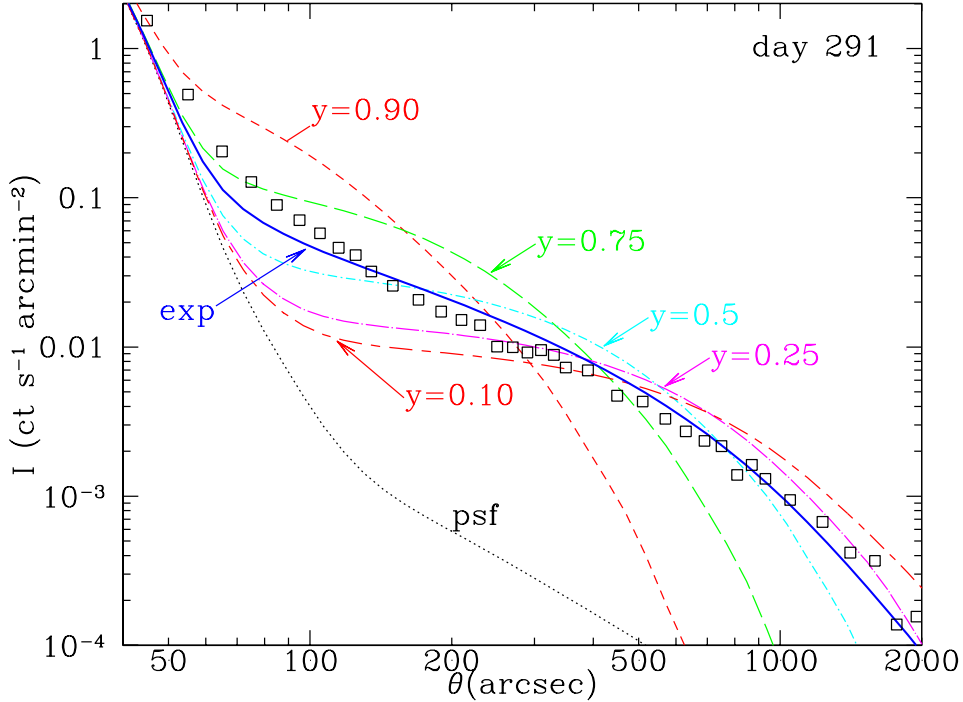}
\end{center}
\vspace*{-2.em}
\caption{
	\label{fig:clump}
	\footnotesize
	Comparison of the X-ray halo produced by 
	smooth exponentially-distributed dust
	with $z_{1/2}=300\pc$ and the nova at $D=2.1\kpc$,
	with models with the same nova distance $D$ 
	and column density $N_{\rm H}$ but in a single
	cloud at $y=r/D=0.1$, 0.25, 0.5, 0.75, and 0.90.
	}
\end{figure}

The profile of the scattering halo depends on the distribution of
dust along the line of sight
(e.g., Predehl \& Klose 1996).
So far we have investigated smooth exponential models.
Other smooth distributions, such as gaussian or $\rm sech^2(z)$
models, do not change the halos significantly, as long as the intervening 
H~I column
remains close to $1.15\times 10^{21}\:{\rm cm^{-2}}$ (this might require the
nova to be at a different distance).
However, since the total reddening to the nova appears to be just 
$E(B-V)\approx0.19$~mag,
it would not be implausible for much of this to be contributed mainly
by one or two diffuse clouds.
In Fig.\ \ref{fig:clump} we show models where the
same column density of dust is located in a single cloud located
at $y=0.10$, $0.25$, $0.5$, $0.75$, or $0.90$ of the distance to the nova.
We see that the halo profile is quite sensitive to the location of the
dust along the line of sight: at $\theta=100$\arcsec, for example,
the halo intensity in the $y=0.10$ case is only 6\% of the
intensity for the $y=0.90$ case.
For a fixed halo angle, there are three separate effects
as the dust cloud is moved from small $y$ to large $y$:
\begin{enumerate}
\item The inverse square law causes the intensity of the radiation
illuminating the dust grains to increase.  This acts to increase the
surface brightness of the dust cloud.
\item The required
scattering angle $\thetascat$ increases [see eq.\ (\ref{eq:scatt_angle}) 
and (\ref{eq:scatt_angle2})].
Since $d\sigma/d\Omega$ tends to decrease for increasing $\thetascat$,
this acts to reduce
the scattered intensity.
\item The time delay $\delta t_1$ increases as $y$ increases.
For $y=0.25$, $\delta t_1= 9.4\days (D/2.0\kpc)(\theta/1000\arcsec)^2$,
while for $y=0.75$,
$\delta t_1 = 84.0\days (D/2.0\kpc)(\theta/1000\arcsec)^2$.
At times when the nova light curve is rising (e.g., day 291) this
leads to weakening of the halo at large angles $\theta$ as $y$ is
increased.
\end{enumerate}
At small halo angles, the near-constancy of $d\sigma/d\Omega$ at
small $\thetascat$ (see Fig.\ \ref{fig:scat}) causes the first effect
to dominate, so the halo intensity increases as $x$ increases
(e.g., at 100\arcsec, the halo is $\sim$5 times brighter for
$y=0.75$ vs.\ 0.25).
At large halo angles, the rapid decline in $d\sigma/d\Omega$ at
large $\thetascat$ also becomes important, as does the time delay if
the lightcurve is rising or falling.

\begin{figure}[h]
\begin{center}
\epsfig{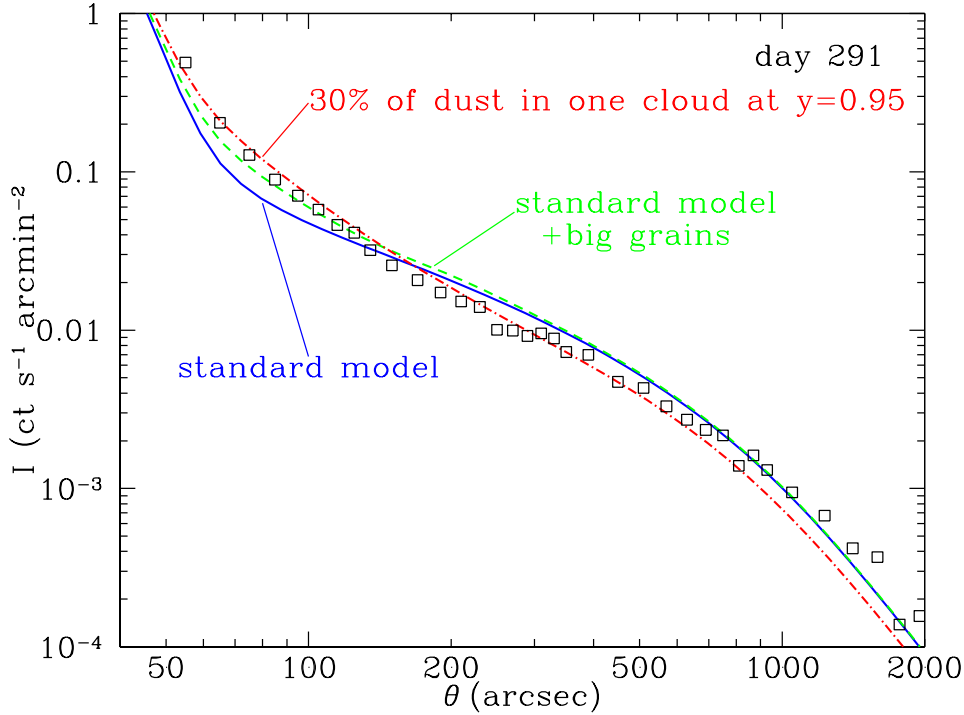}
\end{center}
\vspace*{-2.em}
\caption{
	\label{fig:biggrains}
	\footnotesize
	Halo calculated for nova at $D=2.1\kpc$ and 
	(1) standard dust in an exponential distribution (solid line)
	(2) standard dust plus additional equal mass in large grains
	with $a=2.0\micron$ in an exponential distribution
	(dashed line)
	(3) standard dust, with 70\% of dust in an exponential distribution
	and 30\% in a cloud at $y=r/D=0.95$.
	}
\end{figure}

In Fig.\ \ref{fig:biggrains} we show a model where the total column
density to the nova is held at $1.15\times10^{21}\cm^{-2}$, but 30\% of
this is concentrated in a single ``cloud'' at $y=0.95$, with the remaining
70\% of the dust in an exponential distribution with $z_{1/2}=300\pc$.
The agreement between model and observation is excellent.
We see that the dust in the cloud near the nova raises the halo intensity
at $\theta \ltsim 150\arcsec$, in accord with observations, and removing
30\% of the dust from the general exponential distribution acts to
lower the intensity at 200--500\arcsec, where the model was previously
somewhat above the observed halo.
We show the above ``cloud + exponential'' model for all epochs in
Fig.\ \ref{fig:delays}.  At each time, agreement with the data is
improved relative to the best smooth model, and is in general
excellent.  Note that this model is not necessarily the optimum
``cloud + exponential'' model.  In addition to raising the inner and
lowering the outer halo, the clumpiness in the distribution imprints
structure into the late-time (declining light curve) halos at larger
angles (see Fig.\ \ref{fig:delays}).

Unfortunately, we have little way of determining the
actual distribution of the gas towards the nova, 
since the nova's location at
$l=89.1^\circ$ renders radial velocities unusable for determinations
of distances within $\sim$ 1 or 2 kpc.
In view of these uncertainties, we conclude that the observed
X-ray scattering halo appears to be consistent, within the uncertainties,
with the WD01 dust model.

\subsection{Very Large Grains?}

From Fig.\ \ref{fig:scat} we see that for the WD01
dust model, the X-ray scattering is dominated by the grains with
radii $0.1\micron \ltsim a \ltsim 0.4\micron$.
Grains larger than $0.4\micron$ contribute less than 20\% of the scattering
even at small scattering angles $\thetascat \ltsim 100$\arcsec, 
and less than 1\% at scattering
angles $\thetascat \gtsim 1000$\arcsec.

A recent paper by Witt et al.\ (2001) has argued that the
observed X-ray halo around Nova Cygni 1992 requires that the dust
grain size distribution include significant mass in large dust grains.
Witt et al.\ calculated the dust scattering halo at $t=291\days$ for a 
monochromatic $h\nu=450\eV$ steady point source.  They assumed
$N_{\rm H}=2.1\times10^{21}\cm^{-2}$, and found that a conventional ``MRN''
($dn/da \propto a^{-3.5}$) size distribution for $a < 0.25\micron$
(Mathis, Rumpl, \& Nordsieck 1977), with the dust distributed
uniformly along the line-of-sight, reproduced the observed halo
intensity at $\theta=100\arcsec$, but overpredicted the halo intensity
by a factor $\sim$2 for 300 -- 800\arcsec;
Witt et al.\ did not consider halo angles larger than 800\arcsec.
In order to suppress the scattering at large angles, while maintaining
the observed halo intensity at small angles, they favored size distributions
with the dust mass shifted into larger grains.
However, the modifications to the grain size distribution
proposed by Witt et al.\
are inconsistent with the average optical and ultraviolet extinction
curve for the Milky Way,\footnote{%
	Witt et al.\ note that this size distribution has
	$R_V\equiv A_V/(A_B-A_V)\approx6.2$, exceeding
	the largest values observed in diffuse regions.
	Dust in diffuse regions typically has 
	$R_V=3.1\pm0.5$ .}
and there is no reason to assume that this
particular sightline has an anomalous extinction law.
It is our contention that the excess scattering at 300--800\arcsec\ found by
Witt et al.\ is due primarily 
to assuming too much dust on the line-of-sight -- our
preferred models have $N_{\rm H}\approx 1.15\times10^{21}\cm^{-2}$, only
$\sim55\%$ of the $2.1\times10^{21}\cm^{-2}$ column assumed by Witt et al.

We have taken our best-fitting 
smooth exponential
model -- which uses the standard WD01 size
distribution which reproduces the standard $R_V=3.1$ extinction curve --
and
modified the grain population by adding
very large silicate and
carbonaceous dust grains. 
These grains are arbitrarily given a radius $a=2\micron$, and abundances
such that the total silicate mass and total carbonaceous mass is doubled:
this trial dust model has 50\% of the dust mass in $a=2.0\micron$ grains.
Of course this dust model now has twice as much dust mass per unit area
toward the nova as the original model.

The results of this model are shown in Fig.\ \ref{fig:biggrains}.
The additional dust grains do increase the strength of the scattered halo,
but only slightly.  
The effect is most noticeable in the range 50--100\arcsec, where the
halo intensity is increased by up to 20\%.
We see, however, that even with this unreasonably large mass in large
grains, this model is not superior to the model discussed in
\S\ref{sec:ddist} where 30\% of the
dust is assumed to be located in a cloud at a distance $\sim 100\pc$ from
the nova.
Adding large grains has negligible effect at large angle halos, whereas
placing a cloud close to the nova reduces the intensity at large
halo angles (since the amount of dust at large distances from the nova
is reduced); this reduction appears to improve agreement with the observations.
We conclude that while we cannot rule out a population of large grains,
these grains would have only a modest effect on the scattered halo, and
the observations do not require their presence.
A more important and plausible effect on the halo comes from the presence of 
structure in the ISM.

\section{Discussion\label{sec:discussion}}

\begin{figure}[h]
\begin{center}
\epsfig{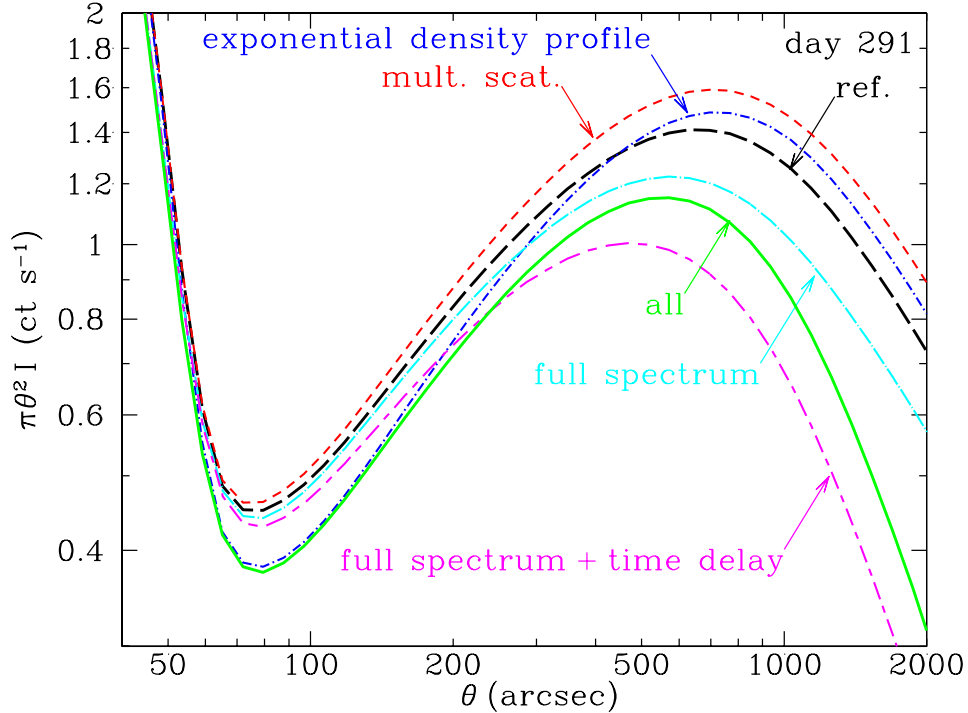}
\end{center}
\vspace*{-2.em}
\caption{
	\label{fig:compare}
	\footnotesize
	$\pi\theta^2I$ for 
	reference model:
	$h\nu=400\eV$, uniformly-distributed dust.
	single scattering only, time delay neglected (curve labelled ``ref'').
	Other curves show effect of adding multiple scattering,
	changing from monochromatic to realistic nova spectrum;
	including effects of time delay for scattered photons, for $D=2.1$kpc;
	and changing from a uniform density to exponential density profile
	with $z_{1/2}=300$pc.
	Curve labelled ``all'' is exponential density model with 
	full spectrum, multiple scattering, and time delay.
	}
\end{figure}

Figure \ref{fig:compare} shows the importance of various features of
the modelling on the intensity of the X-ray halo.
The curve labelled ``ref'' is the single-scattered intensity for a model
where the source is radiating steadily at a single energy $h\nu=400\eV$,
with uniform dust density between observer and source.
Other curves show the effect of including multiple scattering,
time delay, and a realistic nova spectrum (same point source count rate), 
and replacing the uniform dust
distribution with an exponential distribution.
\begin{itemize}
\item Doubly-scattered photons add $\sim$10\% to the intensity 
at $\theta\approx600\arcsec$, and $\sim$20\% at
$\theta\gtsim1200$\arcsec.
\item Changing from a uniform to an exponential dust distribution reduces
the intensity for $\theta\ltsim450$\arcsec\ (less dust close to the source), 
and increases the intensity
for $\theta\gtsim 450$\arcsec\ (more dust far from the source).
The increase is $\sim$10\% for $\theta>800$\arcsec.
\item Replacing the monochromatic spectrum with a realistic spectrum 
reduces the flux at large angles; the adopted $h\nu=400\eV$ is probably
slightly low compared to the more realistic 
spectrum (see Fig.\ \ref{fig:compare_spec}),
resulting in slightly more large angle scattering than for the more
realistic spectrum.
\item Since the light curve is rising at day 291, including time delay
leads to a reduction in the intensity, particularly for larger halo
angles for which the time delays are larger.
At 1000\arcsec, inclusion of time delay reduces the intensity by
$\sim$30\%.
\end{itemize}
Some of these effects increase, others decrease, the intensity.
When all are included together, the intensity for day 291 is reduced at angles
$>60$\arcsec, with a reduction by $\sim$30\% at 1000\arcsec.

For smoothly-distributed dust
and a standard dust-to-gas ratio,
our modelling strongly favors
a column density $N_{\rm H}^{\rm ISM}\approx 1.15\times10^{21}\cm^{-2}$
between us and the nova; this is only
$\sim$50\% of the estimated total gas column in this direction.
This means that the nova must be close enough to have $\sim$50\% of
the column density beyond it.
For the exponential distribution (eq. \ref{eq:exp}), 
this places it at a distance
$\sim2.1(z_{1/2}/300\pc)\kpc$.
Since the actual variation of gas density as a function of
height is not well known, and since 
much of the gas and dust is likely to be contributed by discrete
clouds, it is possible for the nova distance to be as large as $2.5\kpc$ and
still have $\sim$50\% of the gas and dust beyond the nova.

Most estimates of the reddening to the nova have been larger than the
value $E(B-V)=0.20$~mag which we favor.
Austin et al.\ took the weighted mean of 5 different methods, and
obtained $E(B-V)=0.36\pm0.04$~mag.
Our model with $E(B-V)=0.20$~mag has good overall agreement with
the observed halo intensities, but an increase in $E(B-V)$ to 0.36
would result in halo intensities almost twice as strong as
observed.
If the reddening were shown to be $E(B-V)\gtsim0.3$ toward
Nova Cygni 1992, this
would rule out the WD01 dust grain model
which we have used here {\it if} the dust is smoothly distributed.
However, we note that the reddening estimated from the Balmer
decrement at late times is consistent with our estimate.

Furthermore, we show that a plausible modification of the spatial distribution
of the dust can produce good agreement between the observed and
calculated halos: if $\sim30\%$ of the dust (i.e., $E(B-V)=0.06$)
is located in a ``cloud''
within 
$\sim 100\pc$ of the nova, the calculated scattering halo
is in good agreement with observation.
Since such a spatial distribution is not improbable, the observed
X-ray halo around Nova Cygni 1992 does not require the existence of
a population of large grains.
Recent observations by \Chandra\ of the scattered halo around
GX 13+1 also do not support an additional population of large grains
(Smith et al.\ 2002). 

While we have shown that the WD01 dust model is generally
consistent with observations
of Nova Cygni 1992, this conclusion would be overturned if the larger estimates
of the reddening prove to be correct.\footnote{%
	X-rays scattered by dust {\it very} close to the nova 
	would be lost in the point spread function.
	At 0.5 keV the median scattering angle $\Theta_m\approx 1000\arcsec$,
	so most of the photons scattered by 
	dust located at $x = 0.97$ 
	would be at halo angles $\theta < 30\arcsec$.
	Thus dust located at $x > 0.97$ would not make an observable
	contribution to the X-ray halo.
	}
Unfortunately,
Nova Cygni 1992 is a less-than-ideal test of a dust model: as we have
seen, there is uncertainty regarding the value of the reddening $E(B-V)$
(i.e., the total amount of dust); in addition, there is uncertainty
regarding the distribution of gas and dust along the sightline
to Nova Cygni 1992 -- it was
plausible to consider that $\sim$30\% of the dust might be contributed by
a cloud $\sim$100pc from the nova.

The ideal way to study scattering by interstellar dust is to employ
an X-ray source well outside the galactic
plane, so that most of the dust contributing to the scattering is
at $y\ltsim 0.1$,
in which case the
halo angle $\theta$ and scattering angle $\Theta_s$ are nearly the
same for single-scattering.
An extragalactic source would be ideal for this purpose.
In this case, uncertainties regarding the precise location of the
dust in the Galactic disk 
are unimportant (i.e., dust at $y=0$ and dust at $y=0.001$
produce virtually identical halos).
We can hope that the great sensitivity of \Chandra\ and \XMM\ may make
feasible observations of such out-of-plane sources.
Alternatively, a source located in the galactic plane at
$15^\circ \ltsim |l| \ltsim 75^\circ$ would allow HI or CO observations, or
optical absorption line spectroscopy, to locate the absorbing material
using galactic rotation.

The higher angular resolution of \Chandra\ may allow observations at
smaller halo angles, and the improved energy resolution will allow
study of the energy spectrum of the scattered halo.


\section{Summary}

The principal results of this paper are as follows:
\begin{enumerate}
\item We model the X-ray spectrum and light curve of Nova Cygni 1992 using
	the two-component model of Balman et al.\ (1998): 
	optically-thin thermal plasma plus 
	a O-Ne enhanced white dwarf atmosphere.
\item We present the formalism for calculating X-ray scattering halos
	including multiple scattering.
	Time delay is treated exactly
	for single scattering and approximately 
	for multiple scattering.
\item We calculate the X-ray halo for Nova Cygni 1992, using the
	WD01 dust model,
	consisting of a mixture of
	amorphous silicate grains and carbonaceous/PAH grains.
	We compare the calculated X-ray halo with
	\ROSAT\ \PSPC\ imaging of the halo plus point source at 9 epochs.
\item If the dust toward the nova is smoothly distributed (we consider
	an exponential density law as an example)
	we find that the WD01 dust model 
	can quantitatively reproduce the observed halo intensity and
	angular profile provided the 
	reddening $E(B-V)\approx0.20$~mag,
	This is consistent with the value of $E(B-V)\approx0.19$~mag 
	inferred from the observed
	H$\alpha$/H$\beta$ emission line ratio at late times.
\item If $E(B-V)\approx0.20$~mag, then
	the good agreement between the halo calculated for the WD01
	grain model and observations of Nova Cygni 1992 contradicts
	previous claims that the dust toward Nova Cygni 1992 had
	to be either highly porous (Mathis et al.\ 1995) or
	include a substantial population of very large ($a\gtsim 1\micron$)
	dust grains (Witt et al.\ 2001).
\item Improved agreement between model and observation
	is obtained if $\sim30\%$
	of the dust is located in a cloud $\sim100\pc$ from the
	nova.
\item The effects of time delays on the scattered halos depend on the
	distance to the source, which thus provides a method for distance
	determination to non-steady sources.
	The time delay of the halo relative to the nova is clearly visible
	at late times when the nova is in decline.
	The use of time-delay to determine the distance to Nova Cygni 1992
	is discussed elsewhere (Draine \& Tan 2003).
\item It is hoped that future X-ray imaging by \Chandra, \XMM, or other
	telescopes will 
	characterize the X-ray scattering halos around point sources where
	the dust column can be reliably estimated, and where there
	is information constraining the distribution of dust along the
	line of sight.
	Extragalactic X-ray point sources are ideal for this purpose.
	As we have shown here, such observations are capable of strongly
	testing models for interstellar dust.
\end{enumerate}
\acknowledgements
We thank D. Finkbeiner, E.B. Jenkins, and T. Totani for helpful discussions,
J. MacDonald for providing white dwarf model atmosphere spectra in 
computer-readable form, and R.H. Lupton for the SM software package.
This work was supported in part by NSF grant AST-9988126,
and in part by NASA grant NAG5-10811. JCT has received
support via a Spitzer-Cotsen Fellowship from the Department of
Astrophysical Sciences and the Society of Fellows in the Liberal Arts
of Princeton University.

\end{document}